\documentclass[journal]{IEEEtran}
\usepackage{nicematrix}
\usepackage{color}
\usepackage{cite}
\usepackage{textcomp}
\usepackage{xcolor}
\usepackage{balance}
\usepackage{graphicx}
\usepackage[english]{babel}
\usepackage{amsthm,amssymb}
\usepackage{amsmath}
\usepackage{subcaption}
\usepackage{mathtools}
\usepackage{tikz}
\usepackage{todonotes}
\usepackage{pgf}
\usepackage{amsfonts}
\usepackage{bm}
\usepackage[bottom]{footmisc}
\usepackage[ruled,linesnumbered]{algorithm2e}
\usepackage[T1]{fontenc}
\usepackage{bbm}
\usepackage{enumerate}

\usepackage{setspace}
\usepackage{array}
\usepackage{multirow}
\usepackage{diagbox}

\usepackage{todonotes}

\newtheorem{theo}{Theorem}

\DeclareMathOperator*{\argmax}{arg\,max}

\allowdisplaybreaks

\def\BibTeX{{\rm B\kern-.05em{\sc i\kern-.025em b}\kern-.08em
    T\kern-.1667em\lower.7ex\hbox{E}\kern-.125emX}}
\begin{document}

\title{Generalized Group Selection Strategies for \\ Self-sustainable RIS-aided Communication}

\author{Authors}
\author{Lakshmikanta Sau, \textit{Member, IEEE},  Priyadarshi Mukherjee, \textit{Senior Member, IEEE}, \\ and Sasthi~C.~Ghosh, \textit{Member, IEEE}
\thanks{L. Sau, and S. C. Ghosh are with the Advanced Computing \& Microelectronics Unit,  Indian Statistical Institute, Kolkata 700108, India. (E-mail: lakshmikanta@ieee.org, sasthi@isical.ac.in).
}
\thanks{P. Mukherjee is with the Department of Electrical Engineering and Computer Science, Indian Institute of Science Education and Research  Bhopal, India. (E-mail: priyadarshi@ieee.org).}\vspace{-6mm}}



\maketitle
\begin{abstract}
  Reconfigurable intelligent surface (RIS) is a cutting-edge communication technology that has been proposed as a viable option for beyond fifth-generation wireless communication networks. This paper investigates various group selection strategies in the context of grouping-based self-sustainable RIS-aided device-to-device (D2D) communication with spatially correlated wireless channels. Specifically, we consider both power splitting (PS) and time switching (TS) configurations, of the self-sustainable RIS to analyze the system performance and propose appropriate bounds on the choice of system parameters. The analysis takes into account a simplified linear energy harvesting (EH) model as well as a practical non-linear EH model. Based on the application requirements, we propose various group selection strategies at the RIS. Notably, each strategy schedules the $k$-th best available group at the RIS based on the end-to-end signal-to-noise ratio (SNR) and also the energy harvested at a particular group of the RIS. Accordingly, by using tools from high order statistics, we derive analytical expressions for the outage probability of each selection strategy. Moreover, by applying the tools from extreme value theory, we also investigate an asymptotic scenario, where the number of groups available for selection at an RIS approaches infinity. The nontrivial insights obtained from this approach is especially beneficial in applications like large intelligent surface-aided wireless communication. Finally, the numerical results demonstrate the importance and benefits of the proposed approaches in terms of  metrics such as the data throughput and the outage (both data and energy) performance.

\end{abstract}
\begin{IEEEkeywords}
    Reconfigurable intelligent surfaces, spatial correlation, order statistics, $k$-th best selection, extreme value theory.
\end{IEEEkeywords}

\section{Introduction}
\noindent Wireless traffic has surged significantly in recent years due to applications such as the Internet of Things (IoT) and enhanced mobile broadband communication; it is projected to escalate by over fivefold between 2023 and 2028 \cite{ericsson}. In order to handle this massive volume of data, a number of technologies, including beamforming, adaptive modulation, and multiple access frameworks, have been developed in recent decades. All of these innovations are united by the idea of cleverly taking advantage of and adjusting to the wireless channel's random fluctuations rather than controlling them. Conversely, a novel technology known as reconfigurable intelligent surfaces (RISs) aims to maximize the channel's utilization by controlling its random nature \cite{risi2}. Basically, RISs are made up of arrays of passive, reconfigurable elements embedded on a flat metasurface that "controls" the channel rather than adjusting to its changing characteristics \cite{risrev}. The bias voltage can be changed to turn these passive components on and off. They can also reflect the incident signal in the desired direction without the use of radio frequency chains. This lowers the cost of implementation and improves the energy efficiency of the system \cite{impl}.
Additionally, a number of RIS variations \cite{rev3_1, rev3_2, rev3_3} have also led to new research avenues. The authors in \cite{rev3_1} study simultaneously transmitting and reflecting RIS-assisted non-orthogonal multiple access schemes by taking into account hardware impairments and imperfect channel state information (CSI). In order to improve the functionality of cell-free networks aided by reconfigurable holographic surfaces, beamforming algorithms are suggested in \cite{rev3_2}. In \cite{rev3_3}, the performance of active RIS-assisted systems is examined and evaluated.

Motivated by this, the intriguing research avenue of RIS-assisted device-to-device (D2D) communication \cite{risi3, new1,new2} has emerged in the recent years. The authors in \cite{risi3} investigate uplink D2D-enabled RIS-aided cellular networks, in which several D2D links share a single spectrum. By taking into account a generalized Nakagami-$m$ fading scenario, \cite{new1} examines the outage and rate performance of a RIS-assisted D2D communication underpinning cellular network. By using the idea of Markov service processes, a delay-constrained RIS-assisted D2D communication scenario's throughput is analyzed in \cite{new2}. Moreover, the usage of high frequency signals, such as millimeter waves (mmWaves), can help with achieving a high data rate in addition to improving system energy efficiency through the use of RISs \cite{mm10}. While mmWaves-based wireless systems address the issue of increased data rate, they also have a number of drawbacks, including significant propagation and high penetration losses. Hence, RIS-assisted mmWave-based D2D networks seem to be the answer in these kinds of situations, particularly when the direct line of sight (LoS) link is weak. Strategic RIS placement is essential to maximize the benefits of such networks since both the devices communicating with each other must have clear LoS links. With a minimal number of RISs, optimal RIS placement \cite{ls_tgcn,opris} can greatly increase the coverage of such networks, effectively lowering the hardware and installation costs. In this context, the work in \cite{dramp} proposes a double-RIS assisted multihop routing scheme for a D2D communication network. The authors in \cite{tcom_sug} investigate the impact of RISs on the aspect of physical-layer security in terms of both secrecy rate and secrecy outage probability.

Note that the conventional RIS fundamentally reflects the incident signal in a desired direction by adjusting its parameters. It is evident that the training time grows in direct proportion to the size of the RIS \cite{grouping}, which may result in a significant feedback overhead and jeopardize the anticipated performance improvements. Moreover, any given D2D pair always communicates with each other for a limited period of time. As a result, using the entire RIS is not energy efficient and may result in unnecessary resource waste \cite{Tnse}. Therefore, the works in \cite{partition3, grouping} suggest grouping strategies where the nearby RIS elements are grouped together into smaller non-overlapping surfaces with identical phase shifts in order to avoid the massive channel estimation overhead in RIS-based systems \cite{channelestimation}. Without any loss of generality, we consider that each group has an equal number of elements and a particular group can serve one service request at a time. Hence, a single grouping-based RIS can cater to multiple service requests at the same time. The work in \cite{channelestimation} proposes grouping-based RIS, but without considering the impact of spatial correlation. Note that, as the inter-element spacing in an RIS decreases, the aspect of spatial correlation becomes more and more crucial. By incorporating this aspect, the authors in \cite{Tnse} propose a grouping-based RIS-aided multihop communication scenario. Note that the aspect of choosing an appropriate group at the RIS is important because there is always a chance that the best group, which serves our objective, is preoccupied or committed to serving another requesting user pair. However, this aspect of looking at group selection strategies in the RIS, by taking into account the aspect of spatial correlation, has not been adequately explored for the D2D communication scenario. 

Besides, an RIS is a ‘nearly’ passive device, meaning it requires a small but nonzero amount of energy for its phase-shifting operations. In this direction, the work in \cite{ZeRis} proposes a self-sustainable RIS configuration, where the RIS can harvest the required energy from the incoming signals. Depending on the energy harvesting (EH) process, the authors propose two separate architectures: power splitting (PS) and time switching (TS). In the PS configuration, the signal power is divided into two streams: one is used for EH at the RIS and the other is used to reflect the signal in the desired direction. On the contrary, in the TS configuration, the incident signal is solely used for EH during specific time intervals and at other times, it reflects the incoming signal. The authors in \cite{tcom} propose optimized transmission strategies for {\it self-sustainable} RIS-aided simultaneous wireless information and power transfer. As a result, this particular class of RISs is ideal for operation with minimal or zero external energy supply. Hence, it is ideal for scenarios where autonomous operation is beneficial, resulting in an efficient and sustainable wireless communication network. Therefore, we investigate the problem of group selection in the context of a grouping-based self-sustainable RIS-aided wireless topology.

In this manuscript, by considering the aspect of spatial correlation at the RIS, we analytically characterize the end-to-end wireless channel and also provide bounds of operation for the various system parameters of concern. Depending on the applications at hand, we propose generalized $k$-th best group selection strategies by using the tools of higher order statistics \cite{Order_stat} and derive analytical expressions for the outage probability of each of them. Finally, by utilizing the tools from extreme value theory (EVT), we analyse the system's performance under an asymptotic scenario appropriate for large intelligent surface (LIS)-aided communication, where the number of groups is large. Therefore, to summarize, the contributions
of this paper are threefold.

\begin{figure}
    \centering
    \includegraphics[width=0.8\linewidth]{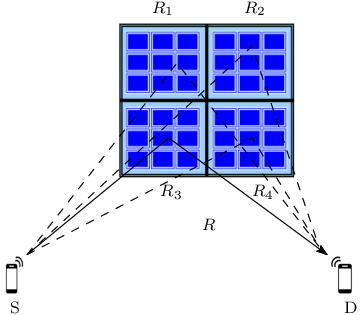}
    \caption{Considered system model.}
    \label{smod}
    \vspace{-6mm}
\end{figure}

\begin{itemize}
\item We consider a grouping-based self-sustainable RIS-aided single hop wireless communication set-up. By taking into account both the PS and TS configurations of the RIS, we propose appropriate bounds on the choice of system parameters. The analysis take into account a simplified linear EH model and also a practical non-linear EH model. Accordingly, a single group at the RIS is selected for information transmission from the source to the destination.

\item By considering the aspect of spatial correlation at the RIS, various group selection strategies are proposed. Specifically, the proposed strategies depend on the end-to-end signal-to-noise ratio (SNR) and the energy harvested at the groups. By employing tools from high order statistics, we propose generalized group selection strategies and also, characterize the outage (both information and energy) probability for each of these schemes. In particular, a complete analytical framework for the performance of the $k$-th best group selection at the RIS is presented. Finally, by using tools from the EVT, we investigate the asymptotic performance of the system, where the number of groups is significantly large.

\item Extensive Monte Carlo simulations validate the proposed analytical framework. We observe that the intelligent group selection strategy outperforms its random selection-based counterpart. Moreover, we compare our method with an existing framework \cite{channelestimation} that groups the RISs without considering spatial correlation, and demonstrate how spatial correlation at the RIS affects the outage performance of the proposed strategies in terms of inter-patch spacing. Furthermore, the impact of the value of $k$ in the proposed generalized $k$-th best group selection strategy can also be seen. The results also depict the effect of having a large number of groups available on the system outage performance. Finally, we also illustrate the importance of the grouping configuration on the system performance.

\end{itemize}
The rest of this paper is organized as follows: System model and problem formulation are explained in Section \ref{sys}, the system characterization is presented in Section \ref{sysc}, and the suggested group selection approaches are discussed in Section \ref{g_sel}. The impact of the extreme value theorem on our proposed scheme is discussed in Section \ref{pevt}, numerical results are presented in Section \ref{simu_res}, and Section \ref{conclu} concludes the work.

\textit{Notations: } The probability distribution function (PDF) and cumulative distribution function (CDF) of a random variable $X$ are denoted as $f_X(x)$ and $F_X(x)$, respectively. $ E[\cdot]$ and $\rm Var(\cdot)$ denote the expectation and variance operators, respectively. Besides, $\Gamma(\cdot)$ and $\gamma(\cdot,\cdot)$ are the complete and the incomplete Gamma function, respectively. $B(\cdot,\cdot)$ is a Beta function, and $\mathcal{I}_\eta(a,b)$ is the incomplete normalized Beta function, which is defined by $\mathcal{I}_\eta(a,b)=\frac{1}{B(a,b)}\int\limits_0^{\eta}y^{a-1}(1-y)^{b-1}$. Furthermore, $I_n(\cdot)$ is the first kind $n$-th order modified Bessel function \cite{grad}.

\section{System Model}\label{sys}
Here, we discuss the considered system model and the related mathematical notations are summarized in Table I.

\subsection{Network Model}\label{net_mod}

As shown in Fig. \ref{smod}, the considered topology consists of a single antenna source $S$ and a single antenna destination $D$. To avoid significant channel estimation overhead, single antenna nodes are more practical for low-power, low-complexity IoT communication scenarios. Here, we assume that, due to the presence of obstacles, the direct LoS link does not exist between $S$ and $D$. Therefore, an RIS, consisting of $N$ reflecting elements, is employed to implement the entire communication process. 


{\color{red}{\begin{table*}[!t]
\centering
 \caption{SUMMARY OF NOTATIONS.} \label{notation}
 \vspace{-2mm}
\resizebox{1.8\columnwidth}{!}{
  \begin{tabular}{|c|c||c|c|}
    \hline \hline
    \textbf{Notation} & \textbf{Description} &  \textbf{Notation} &  \textbf{Description}\\
    \hline
     $N,B$ & Number of RISs and groups respectively & $T_s$ & Slot duration\\
    \hline
     $M$ & Number of elements in each RIS group & $R_i$ & $i$-th RIS \\
    \hline
   $\theta_n$ & Phase shift at $n$-th RIS element & $\lambda$ & Wave length \\
    \hline
    $\rho_L$ & Path-loss at one meter distance & $\alpha$ & Path-loss exponent\\
    \hline
    $d_{m,n}$ & Distance between nodes $m$ and $n$ & $h_i,g_i$ & Complex channel gain \\
    \hline
    $P_{\rm tx}$, $P_{ t},$ and $P_{ph}$ & Transmit , phase shift, and processing power respectively & $h_c,g_c$ & Composite channels\\
    \hline
	$E_{\rm req,PS/TS}$ & Required energy for PS/TS configuration & $\gamma_{th}$ & Required SNR\\ 
    \hline
     $\mathcal{O},\; \mathbb{O}$ & Data and energy Outage probability, respectively. & $E_{l/nl}$ & Harvested energy \\
    \hline
    $R_{\rm req}$ & Required data rate & $E_{\rm req}$ & Required energy \\
    \hline
    $\gamma_{PS/TS}$ & Signal to noise ratio in PS/TS configuration  & $K$ & Rician fading factor\\
    \hline
    $\delta, \beta$ & Scaling and shape parameter & $R_{\rm PS/TS}$ & Achievable data rate \\
    \hline
     $\rho,\zeta$ & Power splitting and time switching factors & $\sigma_0^2$ & Variance of the AWGN\\
    \hline
    $a,b,c$ & Circuit specific constant & $\mathbf{R}$ & Correlation matrix\\
    \hline
  \end{tabular}
  }
  \vspace{-4mm}
\end{table*}}}

\subsection{RIS Characterization}\label{ris_arc}

As stated previously, we consider an RIS consisting of $N$ reflecting elements. These reflecting elements are effectively controlled to adjust both the amplitude and phase of the incident waveform. However, we set the amplitude component to unity for mathematical tractability and simplicity. Moreover, we employ an RIS grouping strategy to reduce the channel estimation cost \cite{partition3}, where the RIS is partitioned into $B$ non-overlapping sub-surfaces  $R_i\; \forall \;i=1,\dots, B$ with $M$ reflecting elements in each, i.e., $N=B\times M$. Note that the values of $M$ and $B$ are determined a-priori, a single group introduces a common phase shift, and at a particular instance, a group can serve only one request \cite{Tnse}. Furthermore, we also assume that the elements of a group are arranged in the form of a uniform planar array (UPA) \cite{upa_lpa}. In this scenario, each $R_i$ has two possible states ON and OFF, and the phase of the incident signal can be changed to a desired direction while it is only in the ON state. Furthermore, a group $R_i$ is configured in such a way, that the energy required for the phase shift operation is obtained by energy harvesting (EH) from the incident signals \cite{ZeRis}. Additionally, we assume that the distance between two adjacent reflecting elements in a group is less than a half-wavelength, i.e., they are spatially correlated. In this context, we consider the following RIS configurations, namely, power splitting (PS) and time switching (TS).

\subsubsection{PS Configuration}  \label{psdef}
Here, the incident signals are designated as two separate streams, namely the information transmission (IT) stream and the EH stream. Specifically, the IT stream requires a sufficient amount of energy, which is provided by the EH stream. Also, the IT and EH stream division is controlled by the tunable power splitting factor  $\rho$ where $0\leq\rho\leq1$. We assume that only $\rho$ portion of the received power is dedicated for EH and the remaining $1-\rho$ portion for IT. Note that, here, EH and IT are simultaneously performed within $R_i$. Since $R_i$ $\forall$ $i=1,...,B$ introduces a common phase shift $\theta_i$, where $\theta_i \in [-\pi,\pi]$, the required amount of energy for information transfer via $R_i$ is 
\begin{equation}\label{E_req_PS}
    {\rm E_{req,PS}}= T_s(KP_t+P_{ph}),
\end{equation}
where $T_s$ is the slot duration, $P_t$ and $P_{ph}$ denote the power consumption of each patch and controller power consumption of $R_i$ for the desired phase shift operation \cite{ZeRis}.


\subsubsection{TS Configuration}  \label{tsdef}
In this architecture, each patch of $R_i$ is entirely devoted to EH or IT during a specific time slot. If $T_s$ is the total time slot and $\zeta\in [0,1]$ fraction of it is required for EH to support the power consumption demand, the entire $R_i$ operates in the EH mode during the interval $[0,\zeta T_s]$ and in the IT mode for the remaining $T_s(1-\zeta)$ time. Hence, in this case, the required amount of energy is given by
\begin{equation}\label{TS_ener}
    {\rm E_{req,TS}}= T_s((1-\zeta)KP_t+P_{ph}).
\end{equation}


\subsection{Channel Model}\label{channel}
As discussed earlier, $S$ communicates with $D$ via an intermediate RIS $R$ that is divided into $B$ non-overlapping surfaces and each surface $R_i$ consists of $M$ reflecting elements. We denote the channel from $S$ to $R_i$ and $R_i$ to $D$  as $\mathbf{h} \in \mathbb{C}^{M\times 1}$ and $\mathbf{g} \in \mathbb{C}^{1\times M}$, respectively. We assume that the wireless links suffer from both large-scale path-loss effect and small-scale block fading. Here, we consider that all the channels $h_j \in \mathbf{h}$ and $g_j \in \mathbf{g}$ $\forall$ $j=1,...,M$ experience Rician fading and their corresponding path-loss factor are $\rho_L d^{-\alpha/2}_{S,j}$ and $\rho_L d^{-\alpha/2}_{j,D}$, respectively, where $\rho_L$ is the path loss at one meter distance, $\alpha$ is the path loss exponent, $d_{S,j}$ and $d_{j,D}$ denote the distance between $S$ to $j$-th element of $R_i$, and $j$-th element of $R_i$ to $D$, respectively. Note that both $|h_j|$ and $|g_j|$ follow the identical distribution \cite{Tnse} as below.
 \begin{align} \label{rice}
 f_{|h/g|}(\alpha,K_{h/g})&=2(1+K_{h/g})e^{-K_{h/g}} \nonumber\\
 & \!\!\!\!\!\!\!\!\!\!\!\!\!\!\!\!\!\!\!\!\!\!\!\!\!\!\!\!\!\!\!\! \!\!\!\!\!\!\!\!\times \alpha e^{-(1+K_{h/g})\alpha^2}I_0\left[ 2\alpha\sqrt{K_{h/g}(1+K_{h/g})}\right] , \:\: \alpha\geq 0.
 \end{align}
 Here $K_{h/g}$ is the Rician factor corresponding to the $S-R_i/R_i-D$ channel, respectively, and $I_0(\cdot)$ denotes the zero-order modified Bessel function of the first kind. 

Since the inter-reflecting element distance of $R_i$ is less than the half-wavelength, we cannot overlook the impact of spatial correlation on the wireless channels \cite{Tnse}. Therefore, the channels from $S$ to $R_i$ and $R_i$ to $D$ are defined as \cite{corr_channel}
\begin{equation}\label{corre}
    \Tilde{\mathbf{h}}=\sqrt{\beta}\mathbf{R}^{1/2} \mathbf{h} \qquad \text{and} \qquad\Tilde{\mathbf{g}}=\sqrt{\beta}\mathbf{g}\mathbf{R}^{1/2} ,
\end{equation}
respectively, where $\beta$ denotes the link gain and $\mathbf{R} \in \mathbb{R}^{M\times M}$ is the spatial correlation matrix. Note that, here we consider the $\it sinc$ model \cite{corr} to characterize $\mathbf{R}_{p,q}$ $\forall$ $p,q=1,\dots,M$, i.e., $\mathbf{R}_{p,q}=\dfrac{\sin \left( \frac{2\pi}{\lambda}d_{p,q} \right)}{\frac{2\pi}{\lambda}d_{p,q}}$, where $\lambda$ is the transmission wavelength and $d_{p,q}$ is the Euclidean distance between the $p$-th and $q$-th element of $R_i$ \cite{corr}.  Accordingly, the composite channel from $S$ to $R_i$ is defined as \cite{partition3}
\begin{equation}\label{com_h}
    h_c=\sum_{i=1}^{M}\Tilde{h_i},
\end{equation}
where $\Tilde{h_i} \in \Tilde{\textbf{h}}$ and $|\Tilde{h_i}|$ follows \eqref{rice}. Similarly, we obtain the composite channel $g_c$ from $R_i$ to $D$ as
\begin{equation}\label{com_g}
    g_c=\sum_{i=1}^{M}\Tilde{g_i}.
\end{equation}

\subsection{Energy Harvesting Model}\label{e_harv}
As stated above, the signal from $S$ reaches $D$ via $R_i$ which is self-sufficient in terms of energy.
Here, every reflecting element of $R_i$ is connected to an RF-to-DC converter, which can draw DC power from the incident signal \cite{ZeRis}.
To explain this phenomenon, several practical EH models have been proposed in the literature. Since the non-linear EH model suggested in \cite{Eharvst} is more theoretically tractable, we consider it in our study. Therefore, the  energy $E_{\rm harv}$, as harvested by a reflecting element of $R_i$ is expressed in terms of the received power $P$ at the reflecting element as
\begin{equation}\label{E_Non}
    E_{nl}=t\Big(\frac{aP|h|^2+b}{P|h|^2+c}-\frac{b}{c}\Big),
\end{equation}
where $|h|^2$ is the power gain of the $S-R_i$ channel, $t$ is the harvesting duration, and $a,b,c$ are the circuit specific parameters. Moreover, as a benchmark, we also consider the linear EH model, i.e.,
\begin{equation}\label{E_L}
    E_{l}= tP|h|^2.
\end{equation}
Note that in \eqref{E_Non} and \eqref{E_L}, we will set the value of $t$ according to the PS and TS configurations, which will be discussed in the following section.

\subsection{Order-Based Selection and Extreme Value Theory}\label{ord_evt}
Assume that $x_i$ with $i \in \{1,\dots,B\}$ represent $B$ i.i.d random variables, which correspond to specific parameters that define the performance of $R_i$. Without loss of generality, we make the following ordering \cite{Order_stat}
\begin{equation}
    x_1\leq x_2 \leq \dots \leq x_B,
\end{equation}
in which training time is used to gain an understanding of this ordering. This ordering is based on the various selection schemes of our proposed framework (will be discussed later). On the basis of these schemes, the $k$-th best group is selected and accordingly, it reflects the incident signals in a desired direction by using its own harvested energy.

Let $i^*$ denote the index of the $k$-th best group for each selection scheme. Therefore, the PDF of $x_i^*$ is given by \cite{Order_stat}
\begin{equation}\label{order}
    f_{x_i^*}(x)=k\binom{B}{k}f_{x_i}(x)F_{x_i}(x)^{B-k}(1-F_{x_i}(x))^{k-1},
\end{equation}
where $F_{x_i}(x)$ and $f_{x_i}(x)$ are the CDF and PDF of $x_i$, respectively.

Furthermore, if $R$ consists of a large number of reflecting elements, i.e., $N \rightarrow \infty$ \cite{LRS},  then we have $B \rightarrow \infty$ for a finite $M$, as $N=M\times B$. In this context, we use the tools of EVT to characterize the system performance. Now, based on the EVT, we get that $x_B$ converges to one of the three limiting distributions: the Gumbel distribution, the Frechet distribution, or the Weibull distribution \cite{Order_stat}. The complete overview of this scenario is analyzed in Section \ref{pevt}.

\section{System Characterization}\label{sysc}
In the considered network topology, $S$ communicates with $D$ via $R_i$, where $R_i$ harvests the energy $E_{\rm req}$ required for the phase shifting operation from the transmitted signal. As stated in Section \ref{ris_arc}, the analytical characterization of this entire process depends on the configuration of the RIS, i.e., PS or TS. Hence, we characterize both the RIS configurations and their impact on the suitable choice of system parameters in the context of the considered network topology.

\subsection{PS Configuration}
According to Section \ref{psdef}, the RIS group $R_i$ harvests energy by using a fraction $\rho$ of the received power, while the remaining is allocated for IT. Therefore, in this case, the received signal at $D$ is
\begin{equation}  \label{ypsdef}
    y_{\rm PS}=\rho_L \sqrt{(1-\rho)P_{\rm tx}(d_{S,R_i}d_{R_i,D})^{-\alpha}}g_ch_c e^{j\phi_i}x+n,
\end{equation}
where $P_{\rm tx}$ is the transmit power, $x$ is the transmitted signal, $\phi_i$ is the associated phase
shift provided at $R_i$, and $n$ is the zero mean additive white Gaussian noise (AWGN), with power $\sigma_0^2$.

Note that, in contrast to the traditional RIS-based method, we do not require a diagonal phase shift matrix of $M$ non-zero elements. Here, the incoming signal receives a common phase shift from the components of $R_i$. Moreover, by rewriting the composite channels $g_c$ and $h_c$ as $|g_c|e^{-\phi_h}$ and $|h_c|e^{-\phi_g}$, respectively, we obtain the optimal phase shift $\phi_{R_i}$ corresponding to $R_i$ as $\phi_{R_i}=\phi_h+\phi_g$ such that the maximum SNR can be achieved. Accordingly, the received signal at $D$ is
\begin{equation}  \label{ypsdefmax}
    y_{\rm PS}=\rho_L \sqrt{(1-\rho)P_{\rm tx}(d_{S,R_i}d_{R_i,D})^{-\alpha}}|g_c||h_c|x+n.
\end{equation}
Next, in the following theorems, we investigate the performance bound on $\rho$, depending on the considered EH model.


\begin{theo}  \label{theo_1}
    By considering a linear EH model and assuming $\rho$ to be equal for all the $M$ reflecting elements of $R_i$, we have
    \begin{equation}  \label{theolin1}
   \frac{MP_{t}+P_{ph}}{\rho_L(d_{S,R_i})^{-\alpha} P_{\rm tx} \sum_{i=1}^{M}|{ h_i}|^2} \leq \rho \leq \frac{\eta}{2^{R_{\rm req}} -1+\eta},
\end{equation}
where $\displaystyle \eta= \frac{\rho_L E_{l,PS} \times d_{R_i,D}^{-\alpha}}{T_s \sigma^2_0 \sum_{i=1}^{M}|{\Tilde{h_i}}|^2 }|g_c|^2|h_c|^2$ and $R_{\rm req}$ is the application-specific minimum required data rate.
\end{theo}

\begin{proof}
See Appendix \ref{thrm1}
\end{proof}

Although the linear EH model is analytically tractable, at times, it fails to capture the actual EH process. For such scenarios, we derive the bound on $\rho$ by considering a nonlinear EH model. From \eqref{E_Non}, for $M$ number of reflecting elements, the total harvested energy  $E_{nl,PS}$ is given by
\begin{equation}\label{E_No1n}
    E_{nl,PS}=T_s\sum_{i=1}^{M}\Big(\frac{a\rho P_{\rm tx} \rho_L(d_{S,R_i})^{-\alpha}|\Tilde{h_i}|^2+b}{\rho P_{\rm tx} \rho_L(d_{S,R_i})^{-\alpha}|\Tilde{h_i}|^2+c}-\frac{b}{c}\Big).
\end{equation} 
Here, to find the bounds for $\rho$, we consider the worst and best channel conditions for EH. This is because, for a given $E_{\rm req,PS}$, the worst channels require a larger $\rho$, while the best channels require a relatively lesser value of $\rho$.
Therefore, with all other parameters remaining constant, we express $E_{nl, PS}$ as a function of the corresponding channel conditions, i.e.,
\begin{equation}
E_{nl, PS}^{\max}=E_{nl, PS} \left( |h_{\rm \max}| \right) \:\: {\rm and} \:\: E_{nl, PS}^{\min}=E_{nl, PS} \left( |h_{\rm \min}| \right),
\end{equation}
where the worst and  best channel conditions $|h_{\rm \min}|$ and $|h_{\rm \max}|$ are defined as 
\begin{equation}  \label{mindef}
    |h_{\rm \min}|= \min\{ |\Tilde{h}_1|^2,\cdots, |\Tilde{h}_M|^2\}.
\end{equation}
and 
\begin{equation}  \label{maxdef}
    |h_{\rm \max}|= \max\{ |\Tilde{h}_1|^2,\cdots, |\Tilde{h}_M|^2\},
\end{equation}
respectively.
Accordingly, in the following theorem, we investigate the performance bound on $\rho$, depending on the considered non-linear EH model.
\begin{theo}\label{th_2}
    By considering a non-linear EH model and assuming $\rho$ to be equal for all the $M$ reflecting elements of $R_i$ in the PS configuration, we obtain
    \begin{align}
   & \frac{c(MP_{t}+P_{ph})}{M\rho_L(d_{S,R_i})^{-\alpha} P_{\rm tx} |h_{\rm \max}|^2\Big(a-\frac{MP_{t}+P_{ph}}{M}-\frac{b}{c}\Big)} \nonumber \\ &  \leq  \rho \leq \frac{\kappa}{(2^{R_{\rm req}}-1)+\kappa} \label{t2},
\end{align}
where $ \displaystyle\kappa=\frac{cE_{nl,PS}^{\min}\rho_L d_{R_i,D}^{-\alpha}}{MT_s |h_{\min}|^2\left(a-\frac{E_{nl,PS}^{\min}}{MT_s}-\frac{b}{c}\right ) \sigma^2_0 } |g_c|^2|h_c|^2.$
\end{theo}
\begin{proof}
    See Appendix \ref{appen2}
\end{proof}

\subsection{ TS Configuration}\label{ts_sec}
In an arbitrary time slot, according to Section \ref{tsdef}, $R_i$ harvests the required energy from the received signal for time $\zeta T_s$ while the remaining time, i.e., $(1-\zeta)T_s$ is dedicated for IT. Therefore, in this case, the received signal at $D$ is 
\begin{align}
y_{\rm TS}=\begin{cases} 
0, & t\leq T_s\zeta\\
\Lambda g_c h_c  e^{j\phi_{i}}x+n, & T_s\zeta < t\leq T_s,
\end{cases}
\end{align}
where $\Lambda=\rho_L\sqrt{P_{\rm tx}}\Big(d_{S,R_i}d_{R_i,D}\Big)^{-\alpha/2}$,  $\phi_i$ is the associated phase
shift at $R_i$, and $n$ is defined earlier in  \eqref{ypsdef}. Accordingly, similar to \eqref{ypsdefmax}, the received signal, corresponding to the maximum SNR that can be achieved, is 
\begin{align}\label{20}
y_{\rm TS}=\begin{cases} 
0, & t\leq T_s\zeta\\
\Lambda |g_c| |h_c|x+n, & T_s\zeta< t\leq T_s.
\end{cases}
\end{align}
Here, we observe that  $\zeta$ has a significant impact both on the EH and IT processes. In this context, we are looking for a compact bound of $\zeta$ that would allow the TS-configuration to perform smoothly. Now, based on our considered EH models, we investigate the performance bound of $\zeta$ for each case independently. In the following theorem, we investigate the performance bound on $\zeta$ for the linear EH model.
\begin{theo}  \label{theo_3}
    By considering a linear EH model and assuming $\zeta$ to be equal for all the $M$ reflecting elements of $R_i$, we have
\begin{equation}\label{th1_3}
   \frac{MP_{t}+P_{ph}}{MP_t+ P_{\rm tx} \sum_{i=1}^{M}|{ h_i}|^2} \leq \zeta \leq 1-\frac{R_{\rm req}}{R_{\rm arc}},
\end{equation}
where $R_{\rm arc}=\log_2({1+\gamma_{\rm TS}})$ is the achievable data rate  and $R_{\rm req}$ is the application-specific minimum required data rate.
\end{theo}

\begin{proof}
See Appendix \ref{appen3}
\end{proof}

Similar to the non-linear EH model-based PS configuration of the RIS, here, we investigate the performance bounds for $\zeta$ by taking into account both the best and worst-case scenarios of the channels. Specifically, we consider the cases $|\Tilde{h_i}|=|h_{\min}|$ and $|\Tilde{h_i}|=|h_{\max}|\; \forall \;i=1,\cdots, M$, where $|h_{\min}|$ and $|h_{\max}|$ are obtained from \eqref{mindef} and \eqref{maxdef}, respectively. In the following theorem, we look at the performance bounds of $\zeta$ for the non-linear EH model.

\begin{theo}  \label{theo_4}
    By considering a non-linear EH model in TS configuration and assuming $\zeta$ to be equal for all the $M$ reflecting elements of $R_i$, we obtain
\begin{equation}\label{ze_n}
   \frac{MP_t+P_{ph}}{M\left(P_t+\frac{a \varphi+b}{ \varphi+c}-\frac{b}{c}\right )}\leq \zeta \leq 1-\frac{R_{\rm req}}{R_{\rm arc}^{\min}}.
\end{equation}
 where $\varphi=\rho_L d_{S,R_i}^{-\alpha} P_{\rm tx}|h_{\max}|^2$ and $R_{\rm arc}^{\min}$ is the achievable data rate for the worst-case scenarios of the channels.
\end{theo}

\begin{proof}
See Appendix \ref{appen4}
\end{proof}

\begin{table*} [!t] 
\begin{center}
  \caption{Summary of results.}
\resizebox{\linewidth}{!}{%
{\renewcommand{\arraystretch}{2} 
  \begin{NiceTabular}{|c||c|c|}[hvlines,cell-space-limits=6pt] 
    \hline
    \bf{Cases} & \bf{Linear EH model} & \bf{Non-linear EH model}\\
    \hline\hline
    PS &  \parbox{7cm}{$\frac{MP_{t}+P_{ph}}{\rho_L(d_{S,R_i})^{-\alpha} P_{\rm tx} \sum_{i=1}^{M}|{ h_i}|^2} \leq \rho \leq \frac{\eta}{2^{R_{\rm req}} -1+\eta},$ \\ where $\eta= \frac{\rho_L E_{l,PS} \times d_{R_i,D}^{-\alpha}}{T_s \sigma^2_0 \sum_{i=1}^{K}|{\Tilde{h_i}}|^2 }|g_c|^2|h_c|^2$. } & \parbox{9cm}{$ \frac{c(MP_{t}+P_{ph})}{M\rho_L(d_{S,R_i})^{-\alpha} P_{\rm tx} |h_{\rm \max}|^2\Big(a-\frac{MP_{t}+P_{ph}}{M}-\frac{b}{c}\Big)} \leq  \rho \leq \frac{\kappa}{(2^{R_{\rm req}}-1)+\kappa},$ \\ where  $\kappa=\frac{cE_{nl,PS}^{\min}\rho_L d_{R_i,D}^{-\alpha}}{MT_s |h_{\min}|^2\left(a-\frac{E_{nl,PS}^{\min}}{MT_s}-\frac{b}{c}\right ) \sigma^2_0 } |g_c|^2|h_c|^2.$} \\
    \hline
    TS & \parbox{5cm}{ $\frac{MP_{t}+P_{ph}}{MP_t+ P_{\rm tx} \sum_{i=1}^{M}|{ h_i}|^2} \leq \zeta \leq 1-\frac{R_{\rm req}}{R_{\rm arc}}.$ }  & $ \frac{MP_t+P_{ph}}{M\left(P_t+\frac{a \varphi+b}{ \varphi+c}-\frac{b}{c}\right )} \zeta \leq 1-\frac{R_{\rm req}}{R_{\rm arc}^{\min}}$, \:\: where $\varphi=\rho_L d_{S,R_i}^{-\alpha} P_{\rm tx}|h_{\max}|^2$. \\
\hline
  \end{NiceTabular}
  }
  }
  \label{tab:summ}
\end{center}
\vspace{-8mm}
\end{table*}

Finally, Table \ref{tab:summ} presents a summary of the main analytical results derived in this section.

\subsection{Composite Channel Characterization} \label{ana_channel}

From Section \ref{channel}, we have the wireless channel from $S$ to the $i$-th element of a group as $h_i \sim \mathcal{CN} \left( m_i,\sigma^2_i\right)$ $\forall$ $i= 1, \cdots, K$. Due to the impact of spatial correlation, from \eqref{corre} we obtain
$\Tilde{\mathbf{h}}=\sqrt{\beta}\mathbf{R}^{1/2}\mathbf{h}$, which can be rewritten as $\Tilde{\mathbf{h}}=\mathbf{R'}\mathbf{h}$, where $\mathbf{R'}=\sqrt{\beta}\mathbf{R}^{1/2}$. In other words, we have
\begin{align} \label{matr}
     &\begin{bmatrix}
        \Tilde{h_1} \\ \Tilde{h_2} \\ \vdots \\ \Tilde{h_M}
    \end{bmatrix} 
=
    \begin{bmatrix}
        r_{11} & r_{12}& \cdots & r_{1M}\\
        r_{21} & r_{22}& \cdots & r_{2M}\\
        \vdots & \vdots &\vdots & \vdots\\
        r_{K1} & r_{K2}& \cdots & r_{MM}  
    \end{bmatrix}
    \begin{bmatrix}
    h_1 \\ h_2\\ \vdots\\ h_M
    \end{bmatrix}.
\end{align}
Hence, from \eqref{matr}, we get $\displaystyle\Tilde{{h_i}}=\sum_{j=1}^M r_{ij}h_j.$
Here, $\Tilde{{h_i}}$ is expressed as a linear combination of $h_j$ $\forall$ $j= 1, \cdots, M$. Therefore, the mean and variance of $\Tilde{h_i}$ are obtained as
\begin{align}
     \mathbb{E}\left[ \Tilde{h_i} \right] &=\mathbb{E}\left[ \sum_{j=1}^M h_jr_{ij} \right]=\sum_{j=1}^M r_{ij}\mathbb{E}[h_j]=\sum_{j=1}^M r_{ij}m_j \:\: {\rm and} \nonumber \\
    \mathrm{Var}\left(\Tilde{h_i}\right)&=\sum_{j=1}^M \mathrm{Var}\left(h_j\right)r^2_{ij}=\sum_{j=1}^M \sigma^2_jr^2_{ij},
 \end{align}
 respectively. Since $m_i$ and $\sigma_i$ $\forall$ $i= 1, \cdots, K$ are finite, by using the Central Limit Theorem \cite{papoulis}, we get
 \begin{equation}\label{cg}
 \Tilde{h_i} \sim \mathcal{CN}\left(\sum_{j=i}^M m_jr_{ij}, \sum_{j=i}^M \sigma^2_jr^2_{ij}\right).
\end{equation}
As a result, by using \eqref{cg} in \eqref{com_h}, we obtain
 \begin{equation}
 h_c=\sum_{i=1}^{M}\Tilde{h_i} \sim \mathcal{CN}\left(\sum_{i=1}^M \sum_{j=1}^M m_jr_{ij}, \sum_{i=1}^M \sum_{j=1}^M \sigma^2_jr^2_{ij}\right).
\end{equation}
Since $h_c$ has a non-zero mean, $|h_c|$ follows the Rician probability density function. Therefore, $|h_c|^2$ follows a non-central $\chi^2$ distribution with the density function \cite{papoulis}
\begin{equation}\label{pdf_chi}
    f_{|h_c|^2}(\mu)=\frac{1}{2}(\frac{\mu}{\Delta_h^2})^{\frac{1}{4}(k_h-2)}e^{-\frac{1}{2}(\mu+\Delta_h^2)}I_{\frac{1}{2}(k_h-2)}(\Delta_h\sqrt{\mu}),
\end{equation}
where $\Delta_h$ is the non-centrality parameter, $k_h$ is the degree of freedom, and $I_n(\cdot)$ is the modified Bessel function of the first kind. Similarly, the density function of $|g_c|^2$ is
 \begin{equation}
    f_{|g_c|^2}(\mu)=\frac{1}{2}\left(\frac{\mu}{\Delta_g^2}\right)^{\frac{1}{4}(k_g-2)}e^{-\frac{1}{2}(\mu+\Delta_g^2)}I_{\frac{1}{2}(k_g-2)}(\Delta_g\sqrt{\mu}).
\end{equation}
From the previous discussion, we know that the received SNR at $D$ is a function of $Z=|g_c|^2|h_c|^2$. However, the product of two non-central $\chi^2$ distributions does not have a closed-form. Hence, we approximate the resulting distribution by the moment matching technique \cite{mom}. We propose using the Gamma distribution for approximation as it is a Type-III Pearson distribution \cite{pearson}, which is widely used in fitting distributions for positive RVs by matching the first and second moments. Specifically, we approximate the product $Z \sim\mathcal{F}(\delta, \beta)$ as Gamma-distributed random variable with the shape and scaling parameters $\delta$ and $\beta$, respectively. Moreover, the PDF and CDF of $Z$ is given as \cite{mom}
\begin{align}\label{jpdf}
    &f_Z(x)=\frac{\beta^-{\delta}}{\Gamma(\delta)}x^{\delta-1}\exp \left(\frac{-x}{\beta}\right), \qquad x>0\;\; \;\text{and}\\
    &F_{Z}(x)=\frac{\gamma\left(\alpha, \frac{x}{\delta }\right)}{\Gamma(\alpha)}, \label{cdf}
\end{align}
respectively, where $\gamma(\alpha,x)$ and $\Gamma(\cdot)$ denote the incomplete and complete Gamma function, respectively.

\section{Proposed Group Selection Strategies}\label{g_sel}
In this section, we propose various group selection strategies for the considered framework, where $R$ is partitioned into $B$ non-overlapping subgroups $R_i$ $i=1,\dots, B$. Specifically, we propose strategies to select one of the $R_i$, corresponding to various performance metrics, as discussed below. Here, the implementation of these strategies is characterized in terms of their respective data and energy outage probabilities. In this context, by using the definition of $\gamma_{\rm PS}$ and $\gamma_{\rm TS}$ from \eqref{12} and \eqref{gts}, respectively, the outage probability, corresponding to a predetermined required data rate $R_{\rm req}$, is defined as
\begin{align}\label{outage3}
    \mathcal{O}_{\rm PS/TS}&=\mathbb{P}\left(\mathbf{f}(\zeta)\log_2(1+\gamma_{\rm PS/TS})<R_{\rm req}\right),
\end{align}
where
\vspace{-3mm}
\begin{align}  \label{fzdef}
\mathbf{f}(\zeta)&=\begin{cases} 
1, & \text{for PS configuration}\\
1-\zeta, & \text{for TS configuration}.
\end{cases}
\end{align}
Similarly, by using \eqref{E_Non} and \eqref{E_L}, the energy outage probability \cite{eout} is defined as
\begin{equation}
    \mathbb{O}_{\rm PS/TS}=\mathbb{P}\left(E_{l/nl}\leq E_{\rm req}\right),
\end{equation}
where $E_{\rm req}$ is a predefined energy threshold. Hereafter,  depending on the outage probability, the group selection mechanisms are described below.

\subsection{Random Group Selection (RGS)}\label{rand}
In this group selection technique, a subgroup is randomly chosen from among those that are accessible to share the information. Here, we define a set $\mathcal{S}$ of all subgroups that can provide the minimum required data rate at $D$, i.e.,
\begin{align}\label{s_def}
    \!\!\!\!S =\left\{R_i\:: \right.& R_{i,{\rm PS/TS}}\geq R_{\rm req}\nonumber\\ & \;\&\;E_{i,{\rm req},{\rm PS/TS}}\leq E_{i,{\rm PS/TS}}, \;\left. i=1,\cdots,B \right\},
\end{align}
where $R_{i,{\rm PS/TS}}$ denotes the achievable data rate at $D$, $E_{i,{\rm req}}$ and $E_i$ are the minimum required and harvested energy corresponding to the $\rm PS$ or $\rm TS$ configuration of $R_i$, respectively. Note that all $R_i \in S$ can harvest the required amount of energy $E_{\rm req}$. Therefore, without loss of generality, if an arbitrary $R_i\in S$ is randomly selected, the resulting outage performance is characterized by the following theorem.
\begin{theo}\label{th_5} 
    The outage probability for the RGS scheme is obtained as
    \vspace{-4mm}
    \begin{align}\label{35}
       &i)\quad \mathcal{O}^{RGS}_{\rm PS} = \frac{\gamma\left(\alpha, \frac{2^{R_{\rm req}}-1}{\delta \big(1-\rho\big)\Psi}\right)}{\Gamma(\alpha)}\nonumber \\
       &ii)\quad \mathcal{O}^{RGS}_{\rm TS} = \frac{\gamma\left(\alpha, \frac{2^{\frac{R_{\rm req}}{1-\zeta}}-1}{\delta \Psi}\right)}{\Gamma(\alpha)}.
    \end{align}
\end{theo}

\begin{proof}
    See Appendix \ref{theo1}
\end{proof}

However, in most practical scenarios, the best subgroup may not be always available. In that case, we have to go for the second-best group selection. Generalizing, we concentrate on how the outage probability for the various selection schemes is affected by the $k$-th best group selection. Accordingly, the outage probability for the $k$-th best group selection scheme is 
\begin{equation}\label{outage}
    \mathcal{O}_{k,\rm PS/TS}=\mathbb{P}\left(\mathbf{f}(\zeta)\log_2(1+\gamma^*_{\rm PS/TS})<R_{\rm req}\right),
\end{equation}
where $\gamma^*_{\rm PS/TS}$ is the received SNR at $D$ for the $k$-th best group. Therefore, the $k$-th best group selection schemes, based on different system parameters, are presented below.

\subsection{SNR-Based Group Selection (SBGS)} 
Here, we investigate a group selection strategy based on the achievable SNR. In this context, we consider the CSI to be known a priori. Specifically, in this scheme, the group that attains the $k$-th highest SNR at $D$ is the $k$-th best group, and we select that one accordingly. Note that all the available groups constitute the set $\mathcal{S}$ as defined in \eqref{s_def}. Consequently, let us consider that $X_1\leq X_2 \leq \cdots \leq X_{|\mathcal{S}|}$ be the ordering of the received SNR and the index of the group corresponding to the $k$-th best SNR is $i^*$. That is,
\begin{equation}
    i^*=\argmax_{i\in \{ 1,\dots,|\mathcal{S}|\}}^{(k)}\{X_1,\dots,X_{|\mathcal{S}|}\}.
\end{equation}
Therefore, from \eqref{order}, we obtain the PDF of $X_{i^*}$ as
\begin{equation}\label{pdf_or}
    f_{X_{i^*}}(x)=k\binom{|\mathcal{S}|}{k}f_{X_i}(x)F_{X_i}(x)^{|\mathcal{S}|-k}(1-F_{X_i}(x))^{k-1},
\end{equation}
where $f_{X_i}(x)$ and $F_{X_i}(x)$ are provided in \eqref{jpdf} and \eqref{cdf}, respectively. In the following, the outage performance for the SBGS scheme is characterized by the following theorem.
\begin{theo}\label{th_6}
    The outage probability for the SBGS scheme, where the $k$-th best group is selected, is determined by
\begin{equation}\label{39}
    \mathcal{O}^{SBGS}_{k,\rm PS/TS}=\mathcal{I}_{F_{X_i}(x)}\left(|\mathcal{S}|-k+1,k\right),
\end{equation}
\end{theo}
\begin{proof}
    See Appendix \ref{theo2}.
\end{proof}


According to the system configuration considered, the residual harvested energy has a significant impact on data communication. Therefore, we will discuss the energy-based group selection scheme in detail below.

\subsection{Energy-Based Group Selection (EBGS)}
In this scheme, each $R_i \; i=1,\dots,B$ harvests the required energy from the incident signals. Note that the EH process depends on the RIS characterization, the considered EH models, and also the correlated wireless channels from $S$ to $R$, i.e., on $\Tilde{h_i},\; i=1,\dots, M$. Here, we consider the linear as well as the nonlinear EH models for both the PS and TS configurations of the considered framework. We define the set of groups that are able to harvest the minimum required energy and support the minimum data rate as  
\begin{align}
    \!\!\!\!\mathcal{A} =\left\{E_i\:: \right. \;\;&E_{i,{\rm req},{\rm PS/TS}}\leq E_{i,{\rm PS/TS}}, \nonumber\\& \& \;R_{i,{\rm PS/TS}}\geq R_{\rm req}\;\left. i=1,\cdots,B \right\},
\end{align}
where $E_i$ is the energy harvested by the $i$-th group. In this scheme, the group that harvests the $k$-th highest energy is defined as the the $k$-th best group, and we select that one accordingly. Consequently, let us consider that $E_{1}\leq E_{2} \leq \cdots \leq E_{|\mathcal{A}|}$ be the ordering of the harvested energy and the index of the group corresponding to the $k$-th best harvested energy is $i^*$, i.e.,
\begin{equation}
    i^*=\argmax_{i\in \{ 1,\dots,|\mathcal{S}|\}}^{(k)}\{E_1,\dots,E_{|\mathcal{A}|}\}.
\end{equation}
Therefore, from \eqref{order}, we obtain the PDF of $E_{i^*}$ as
\begin{equation}\label{pdf_en}
    f_{E_{i^*}}(x)=k\binom{|\mathcal{A}|}{k}f_{E_i}(x)F_{E_i}(x)^{|\mathcal{A}|-k}(1-F_{E_i}(x))^{k-1},
\end{equation}
where $f_{E_i}(x)$ and $F_{E_i}(x)$ are the PDF and CDF of $E_i$, respectively. Accordingly, the energy outage for the proposed EBGS scheme is characterized by the following theorem.
\begin{theo}\label{th_7}
    The energy outage probability for the EBGS scheme in both the PS and TS configurations for the $k$-th best group is evaluated as
\begin{equation}\label{49}
    \mathbb{O}^{EBGS}_{k,\rm PS/TS}=\mathcal{I}_{F_{E_i}(x)}\left(|\mathcal{A}|-k+1,k\right).
\end{equation}
\end{theo}
 
\begin{proof}
    See Appendix \ref{theo3}
\end{proof}

\noindent \textbf{Remark:} Note that, both SBGS and EBGS consider a perfect CSI scenario. We  assume that, for all $i$, the $S-R_i$ and $R_i-D$ channels remain constant during a channel coherence block of length $T_c$ and that this satisfies $T_p<T_c$, where $T_p$ is the pilot length \cite{channel_es}. Then the estimate for the channels is obtained in \cite{opt_channel}, which exists iff $T_p \geq M+1$. This very condition implies that the channel estimation overhead $T_p$ scales at least linearly with $M$ \cite{partition3}. Moreover, by considering alternative assumptions, such as models involving sparsity, more efficient channel estimation methods may also be employed. In such cases, the overhead may be even lesser. Furthermore, the work in \cite{imperfect_chnl} investigates the aspect of imperfect wireless channel estimation for RIS-aided systems. Therefore, based on the framework proposed here and by using \cite{imperfect_chnl}, we can extend the present work and propose an intelligent channel estimation strategy-based generalized group selection framework for a self-sustainable RIS-aided set-up.



\section{EVT-Based Performance Analysis}\label{pevt}
Here, we investigate the asymptotic performance of the proposed selection strategies in terms of the number of groups, i.e., when $B\rightarrow \infty$. Specifically, to analyze the performance of the proposed $k$-th best selection schemes, we derive the limiting distribution of the $k$-th best group and use EVT tools to assess the asymptotic outage probability. In this context, as mentioned in Section \ref{ord_evt}, let us assume that $X_B$ denotes the largest order statistic of $B$ i.i.d. random variables and the corresponding PDF and CDF are denoted as $F_{X_B}(x)$ and $f_{X_B}(x)$, respectively. Note that the random variable of interest for the EBGS scheme in PS configuration corresponding to the linear and nonlinear EH model is obtained by using \eqref{abc1} and \eqref{E_No1n}, respectively. Similarly, \eqref{12} and \eqref{12a} are used to obtain the random variable of interest for the SBGS scheme in PS configuration corresponding to the linear and nonlinear EH model, respectively. Accordingly, the value of $\rho$ and $\zeta$ will be set to obtain the random variables in the TS configuration as stated earlier in Section \ref{ts_sec}.

\begin{figure*}[h!]
 \begin{subfigure}[b]{.5\textwidth}
    \centering
    \includegraphics[width=0.62\linewidth]{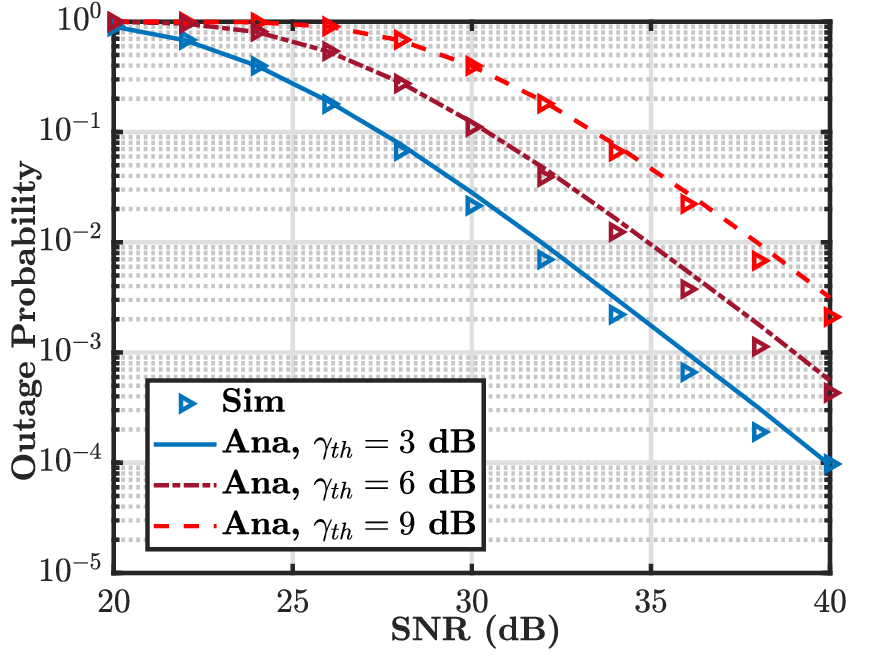}
    \vspace{-2mm}
    \caption{}
    \vspace{-2mm}
    \label{cov}
\end{subfigure}
\begin{subfigure}[b]{.5\textwidth}
    \centering
    \includegraphics[width=0.62\linewidth]{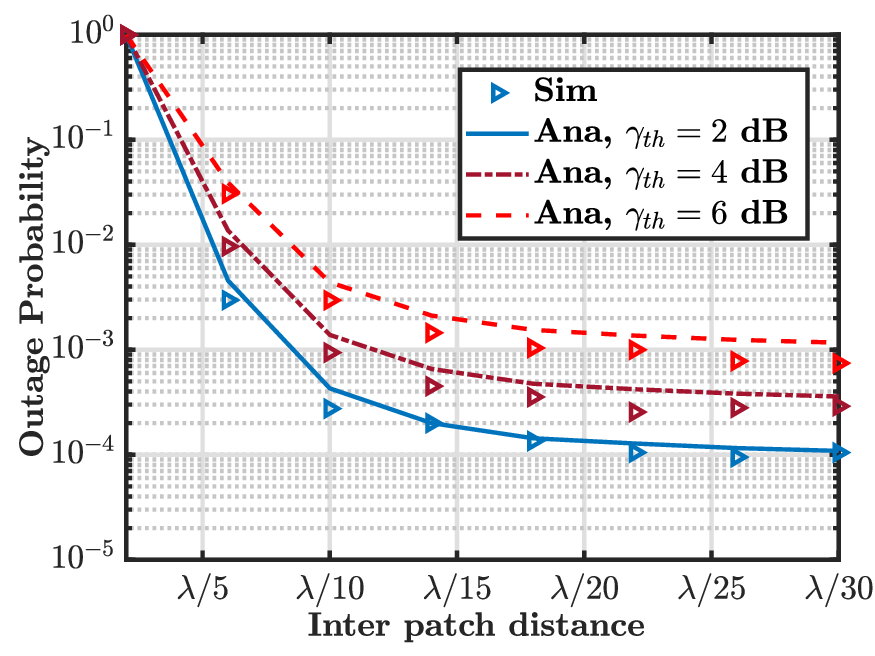}
    \vspace{-2mm}
    \caption{}
    \vspace{-2mm}
    \label{div}
\end{subfigure}
\caption{\footnotesize  Validation of concept for different SNR thresholds. Variation of outage with (a) SNR, and (b) inter-patch distance.}
\vspace{-4mm}
\end{figure*}

Now we consider that $H(x)$ is the limiting CDF of $\frac{X_B-\kappa}{\tau}$, where $\kappa$ and $\tau$ are normalizing constants. In our proposed group selection schemes, it can be shown that
\begin{equation}
    \lim_{x\rightarrow \infty} \frac{1-F_{X_i}(x)}{f_{X_i}(x)}=l, \; 0<l \; \forall\;i=1, \dots B.
\end{equation}
As a result, the limiting distribution of the $k$-th best group follows the Gumbel distribution \cite{david_order} with CDF
\begin{equation}\label{g}
    H(x)=\exp(-\exp(-x)), \; -\infty <x< \infty.
\end{equation}
Therefore, for a large $B$, the limiting CDF of the $k$-th best group is defined as \cite{david_order}
\begin{align}\label{gk}
    H^{(k)}(x)&=\frac{1}{(k-1)!}\int_{\psi(x)}^\infty e^{-t}t^{k-1}dt=H(x)\sum_{j=0}^{k-1}\frac{\left(\Psi(x)\right)^j}{j!},
\end{align}
where $\Psi(x)=-\log H(x)$. By using \eqref{g} and \eqref{gk}, we obtain
\begin{equation}
    H^{(k)}(x)=\exp\left(-\exp\left(-x\right)\right)\sum_{j=0}^{k-1}\frac{\exp\left(-jx\right)}{j!}.
\end{equation}
Hence, the outage probability for the $k$-th best group is
\begin{align}
    &\mathcal{O}^{EVT}_{k}(x)=P\{X_{B-k+1}\leq x\}\nonumber\\
    & =P\left\{\frac{X_{B-k+1}-\kappa}{\tau}\leq \frac{x-\kappa}{\tau}\right \}=H^k\left(\frac{x-\kappa}{\tau}\right).
\end{align}
Next we discuss the impact of EVT-based performance on both the SBGS and EBGS schemes. In the SBGS (EBGS) system, the asymptotic outage probability of the $k$-th best group selection is expressed as
\vspace{-2mm}
\begin{equation}\label{59}
    \mathcal{O}^{EVT}_{k}(x)=\exp\left(-\exp\left(-\frac{x-\kappa_{l}} {\tau_{l}}\right)\right)\sum_{j=0}^{k-1}\frac{\exp\left(-j\frac{x-\kappa_{l}}{\tau_{l}}\right)}{j!},
\end{equation}
where $l\in \{SBGS,EBGS\}$ and the corresponding constants $\kappa_{l}$ and $\tau_{l}$ are derived from the PDF of the $k$-th best group selection scheme according to the SBGS (EBGS) scheme.
\begin{figure*}[h!]
 \begin{subfigure}[b]{.5\textwidth}
    \centering
    \includegraphics[width=0.64\linewidth]{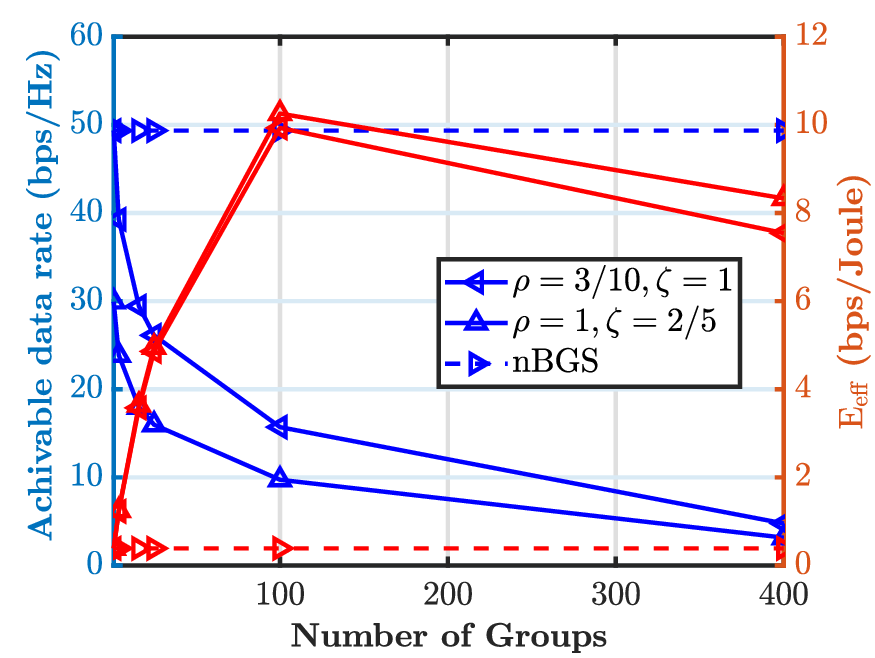}
    \vspace{-2mm}
    \caption{}
    \vspace{-2mm}
    \label{ee}
\end{subfigure}
\begin{subfigure}[b]{.5\textwidth}
    \centering
    \includegraphics[width=0.62\linewidth]{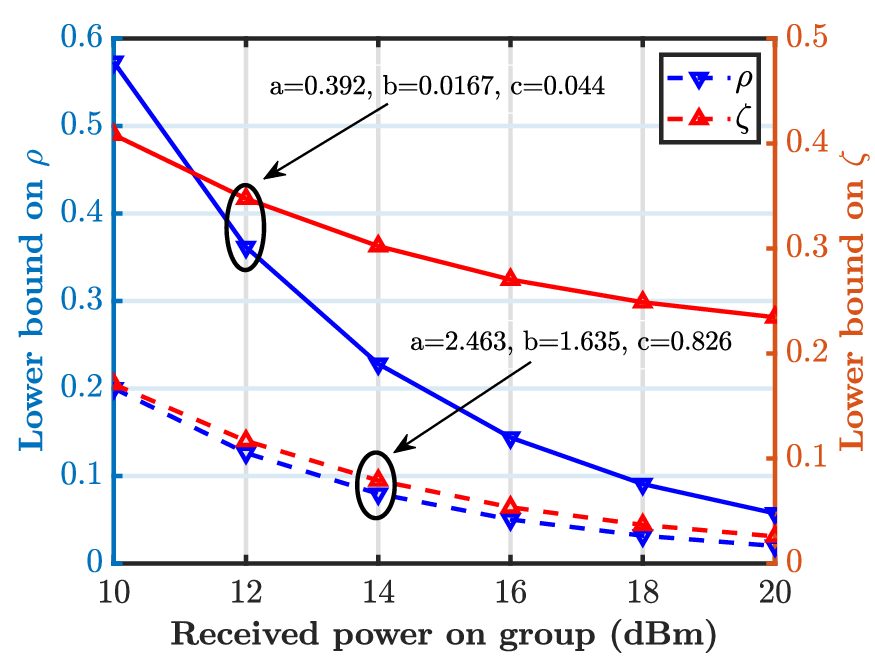}
    \vspace{-2mm}
    \caption{}
    \vspace{-2mm}
    \label{rhobound}
\end{subfigure}
\caption{\footnotesize  (a) Performance trade-off investigation, and (b) Impact of received power on lower bound.}
\vspace{-4mm}
\end{figure*}

\section{Simulation Results}\label{simu_res}
In this section, we now validate the derived analytical results. We consider a Rician fading scenario with the following system parameters: transmission power $P_{\rm tx}=30$ dBm \cite{dramp}, path-loss factor $\rho_L$ at a distance of one meter is $\rho_L=10^{-3.53}$, path-loss exponent $\alpha=2$,  slot duration $T_s=100\; \mu s$ \cite{Tnse}, wavelength $\lambda=0.1$ m \cite{dramp}, and noise power $\sigma^2_0=-105$ dBm \cite{bandw}. Moreover, we consider that the total number of patches of an RIS is $400$, i.e., $N=400$, and the inter-patch distance is $\lambda/8$ \cite{Tnse}. Furthermore, the normalizing constants for the considered non-linear EH model are $a = 2.463,\; b = 1.635,$ and $c = 0.826$ \cite{nparaf}. Finally, we conclude by comparing our proposed approach with an existing benchmark scheme.


Fig. \ref{cov} compares the analytically obtained outage probability in Section \ref{g_sel} with the extensive Monte Carlo simulations performed. For this, we consider that $R_i$ $\forall$ $i=1,\cdots,20$ consists of $20$ patches, the inter-patch distance is $\lambda/8$, and the Rician fading factor $K=1$. Here, by considering the correlated channels, we perform the procedure $10^6$ times in the Monte Carlo simulation, figuring out the optimal phase shift each time. Therefore, we observe from the figure that, irrespective of the SNR threshold $\gamma_{th}$, the outage probability exhibits a decreasing trend with respect to the SNR, which is the function of distance, path-loss, transmit power, and noise. This is quite intuitive because the data rate increases with the SNR, leading to a downward trend in the outage probability. Moreover, for a particular SNR, we also observe an increased outage probability with increasing $\gamma_{th}$; for example, observe the performance gap at ${\rm SNR}=30$ dB between $\gamma_{th}=3$ dB, $\gamma_{th}=6$ dB, and $\gamma_{th}=9$ dB. It can be observed that the analytical results closely resemble the simulation results, which validates our analytical framework.

\begin{figure*}[t!]
\begin{subfigure}[b]{.32\textwidth}
    \centering
    \includegraphics[width=0.92\linewidth]{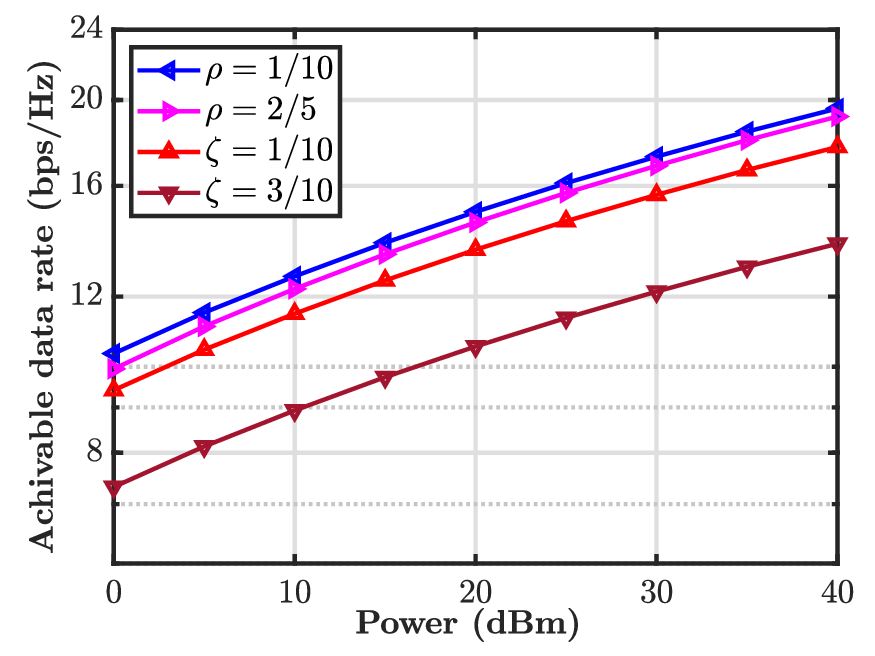}
    \vspace{-2mm}
     \caption{}
    \vspace{-2mm}
    \label{psts}
\end{subfigure}
 \begin{subfigure}[b]{.32\textwidth}
    \centering
    \includegraphics[width=0.92\linewidth]{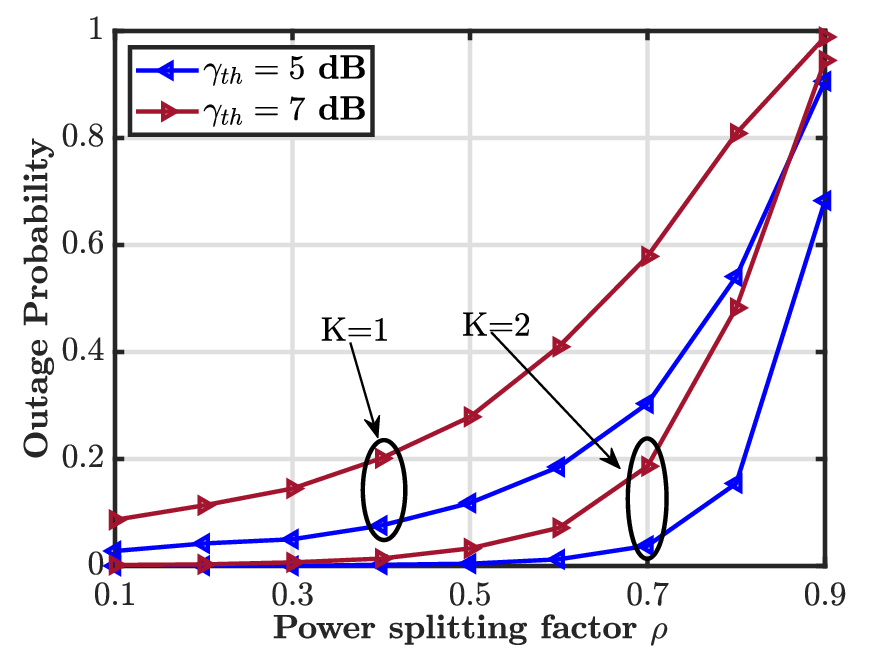}
    \vspace{-2mm}
    \caption{}
    \vspace{-2mm}
    \label{rho_pt}
\end{subfigure}
\begin{subfigure}[b]{.32\textwidth}
    \centering
    \includegraphics[width=0.92\linewidth]{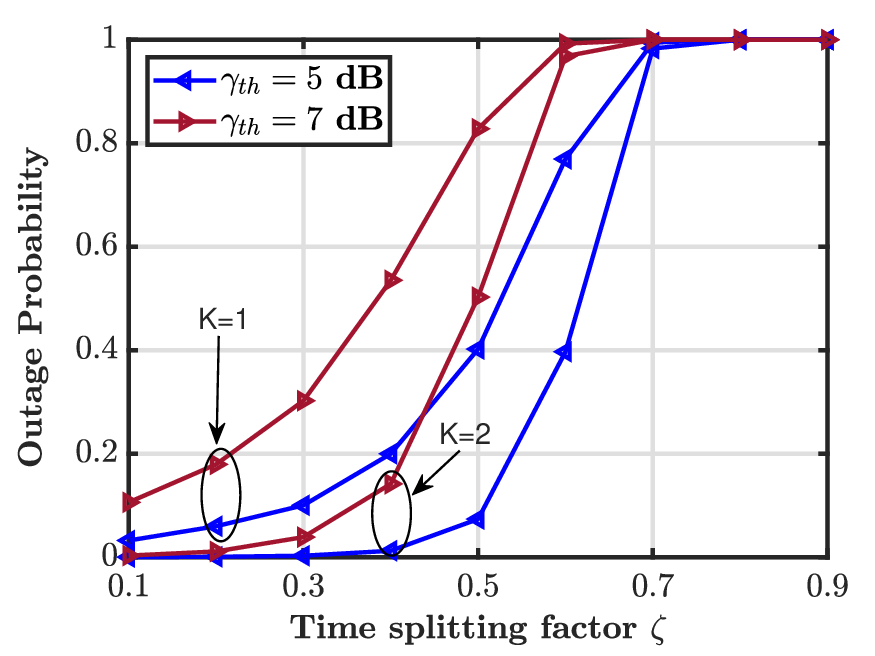}
    \vspace{-2mm}
     \caption{}
    \vspace{-2mm}
    \label{zeta_out}
\end{subfigure}

\caption{\footnotesize  Impact of (a) $\rho$ and $\zeta$ on data rate. Variation of (b) $\rho$, and (c) $\zeta$ with outage performance for different Rician values.}
\vspace{-4mm}
\end{figure*}

With identical system parameters as in Fig. \ref{cov} and also considering a transmit power $P_{\rm tx}=30$ dBm, Fig. \ref{div} investigates the impact of the inter-patch distance on the outage probability. Thereafter, we match the analytical results obtained with the corresponding Monte Carlo simulation, which further verifies our analytical framework. We observe that the outage probability for $\gamma_{th}=2$ dB, $\gamma_{th}=4$ dB, and $\gamma_{th}=6$ dB exhibits a decreasing trend with the decreasing inter-patch distance. This is because a decreasing inter-patch distance enhances the channel correlation, which results in more gain on the received SNR. This also corroborates the observation made in \cite{Tnse}. Moreover, as observed in Fig. \ref{cov}, here also we notice that the outage probability corresponding to a smaller SNR threshold is less than its higher counterpart.

\begin{figure*}[h!]
\begin{subfigure}[b]{.32\textwidth}
    \centering
    \includegraphics[width=0.92\linewidth]{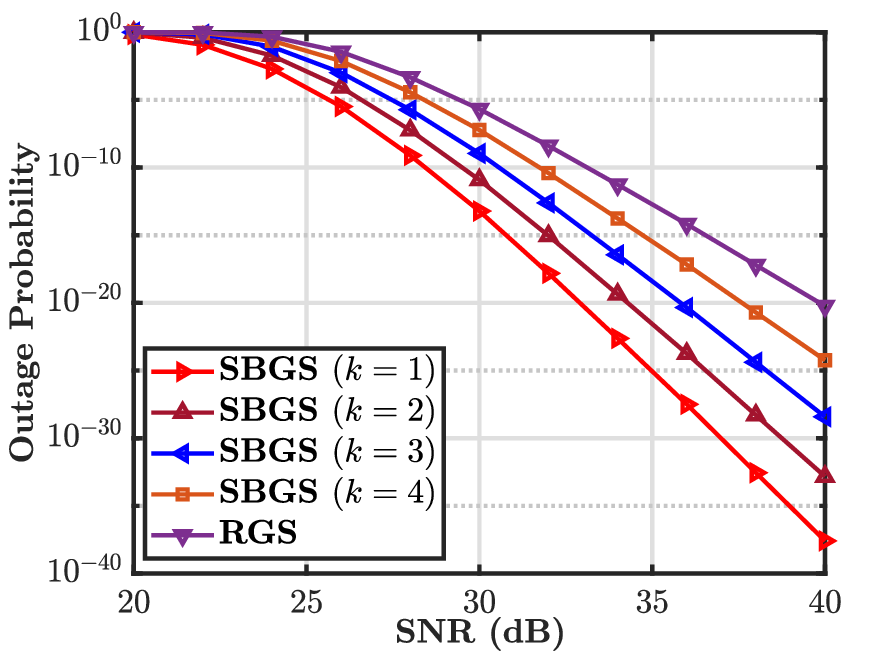}
    \vspace{-2mm}
     \caption{}
    \vspace{-2mm}
    \label{sbgs}
\end{subfigure}
\begin{subfigure}[b]{.32\textwidth}
    \centering
    \includegraphics[width=0.92\linewidth]{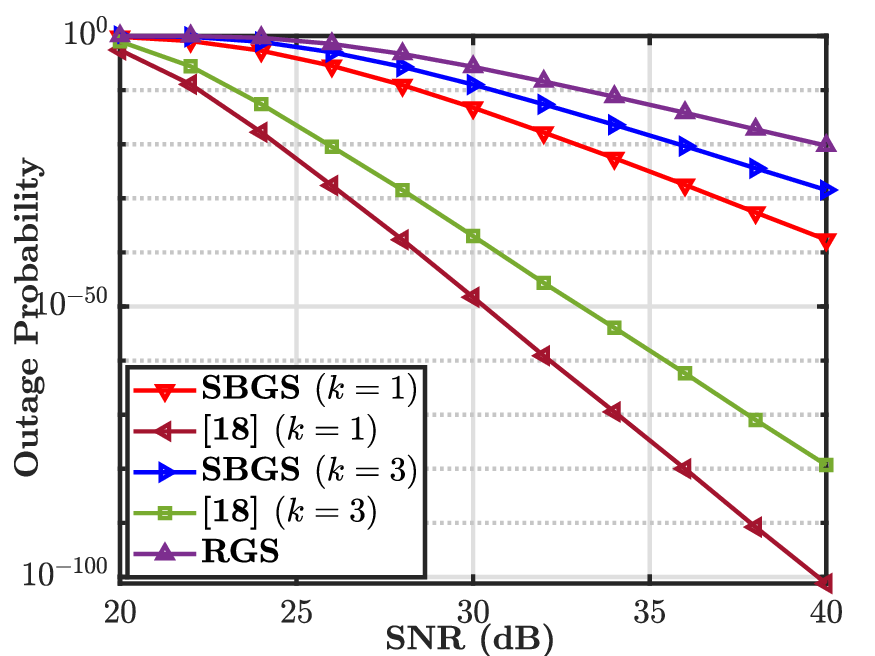}
    \vspace{-2mm}
    \caption{}
    \vspace{-2mm}
    \label{snr_comp}
\end{subfigure}
\begin{subfigure}[b]{.32\textwidth}
    \centering
    \includegraphics[width=0.92\linewidth]{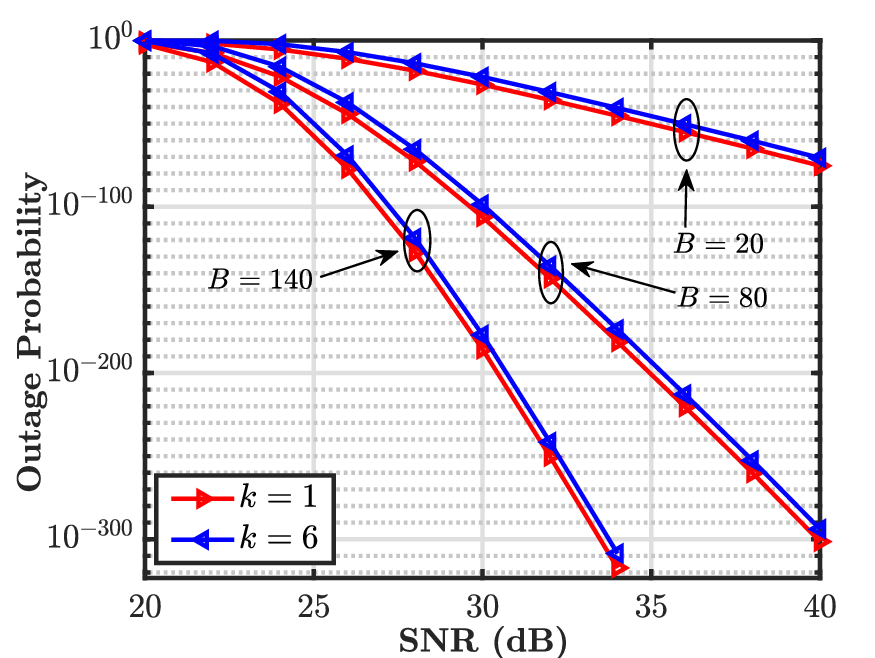}
    \vspace{-2mm}
    \caption{}
    \vspace{-2mm}
    \label{evt}
\end{subfigure}
\caption{\footnotesize  (a) Impact of the SBGS's $k$-th selection, (b) Comparison of the proposed and an existing scheme, and (c) Impact of EVT in SBGS.}
\vspace{-4mm}
\end{figure*}

Based on \eqref{E_req_PS}, \eqref{TS_ener}, and the discussion in Section \ref{sysc}, the energy-efficiency of a particular group is defined as 
\begin{equation*}
   \rm E_{eff}= \frac{R_{\rm arc}}{E_{\rm req,o}} \quad {\rm o} \in \{ {PS,TS} \},
\end{equation*} 
where $R_{\rm arc}$, and $E_{\rm req,o}$ are the achievable data rate and energy consumption of a group, respectively. Fig. \ref{ee} illustrates the importance of RIS grouping on the achievable data rate and {$\rm E_{eﬀ} $}, as a function of the number of groups of a RIS, which consists of $400$ patches. Specifically, we investigate the consequences of adopting a grouping-based scenario (GBS) and a no grouping-based scenario (nGBS). The effect of both the PS and TS configurations in GBS is investigated by considering  $\left(\rho=3/10,\zeta=1\right)$ and $\left(\rho=1,\zeta=2/5\right)$, respectively. We observe that, in both cases, the achievable data rate exhibits a downward trend with increasing number of groups. This is because an increasing number of groups implies a decreasing number of elements per group, which results in a lower data rate. 
Conversely, it is interesting to observe that $\rm{E_{eff}}$ increases up to a certain group size and then decreases thereafter. The reason for this is attributed to the fact that, $\rm{E_{eff}}$ is defined as the ratio of $R_{\mathrm{arc}}$, which is a logarithmic function of a positive quantity, to the required energy, $E_{\rm{req,o}}$, which is a linear function of the same. Consequently, beyond a certain point, the linear growth in energy consumption dominates, leading to a reduction in $\rm{E_{eff}}$. On the contrary, as the entire RIS is being used in nGBS and the required number of phase shifts equals to the total number of reflecting elements, the aspect of grouping does not have any impact on the achievable data rate and $\rm{E_{eff}}$. Hence, we suggest using the GBS framework depending on the application, such as in a dense environment, which improves $\rm E_{eff}$ performance and allows the RISs to serve more users at a time.

Fig. \ref{rhobound} shows the impact of the considered nonlinear EH model parameters on the derived analytical bounds for $\rho$ and $\zeta$ corresponding to both the PS and TS configurations of the considered RIS. In this case, we consider the Rician factor $K=2$, a group consists of $100$ patches, and the received power is a function of transmit power, path-loss factor, and the distance from the $S$ to the group. We obtain the lower bound for both the PS and TS configurations by utilizing two distinct sets of parameters $\{a=2.463, b=1.635, c=0.826\}$ and $\{a=0.392, b=0.0167, c=0.044\}$ in \eqref{t2} and \eqref{ze_n} as suggested in \cite{Eharvst}. Irrespective of the parameters, the figure demonstrates an overall decreasing trend of both the lower bounds of $\rho$ and $\zeta$ with the received power in a group. The reason for this observation is justified by the fact that, for a fixed energy requirement, an increase in the received power leads to a decrease in both $\rho$ and $\zeta$.

Fig. \ref{psts}, for both the PS and TS configurations, demonstrates an overall increasing trend of the achievable data rate with increasing transmitted power. Here, to obtain the achievable data rate from \eqref{ypsdef} and \eqref{20}, we consider $K=1$, the distance from $S$ to $R_i$ and $R_i$ to $D$ is $15$ m and $20$ m, respectively. It is observed that the achievable data rate for a smaller $\rho$  $(\zeta)$,  outperforms the performance for a higher value of $\rho$  $(\zeta)$. This is because, according to the PS configuration, the higher $\rho$ helps to harvest more energy at the RIS, which leads to a lower data rate at $D$. In contrast, for a smaller $\zeta$, the energy harvesting time is small, resulting in a higher data rate at $D$. Note that in both the PS and TS configurations, we choose the value of $\rho$ and $\zeta$ in such a way that they are able to harvest the minimum required energy for IT as well as satisfy the minimum data required conditions as shown in Table \ref{tab:summ}.

We observe that the outage probability exhibits an increasing trend with increasing $\rho$ and $\zeta$ in Fig. \ref{rho_pt} and Fig. \ref{zeta_out}, respectively. In this case, simulations are performed for two distinct Rician factors $K=1$ and $K=2$, as well as two different SNR thresholds $\gamma_{th}=5$ dB and $\gamma_{th}=7$ dB. Here, the other system parameters are considered as in Fig. \ref{cov}. Fig. \ref{rho_pt} depicts that the outage probability as obtained in \eqref{outage3}, in PS configuration, for all predefined SNR thresholds, exhibits an increasing trend with increasing power splitting factor. This is because, as mentioned earlier, a higher $\rho$ aids in harvesting of more energy, which lowers the data rate and causes more outages. Similarly, for TS configuration, Fig. \ref{zeta_out} shows that the data outage exhibits an increasing trend with increasing $\zeta$. In both cases, we also observe that for a particular $K$, the outage probability for $\gamma_{th}=7$ dB is more than $\gamma_{th}=5$ dB. Moreover, as a higher $K$ implies a higher data rate, the outage probability for $K=1$ outperforms the outage probability for $K=2$. 

\begin{figure*}[h!]
\begin{subfigure}[b]{.5\textwidth}
    \centering
    \includegraphics[width=0.62\linewidth]{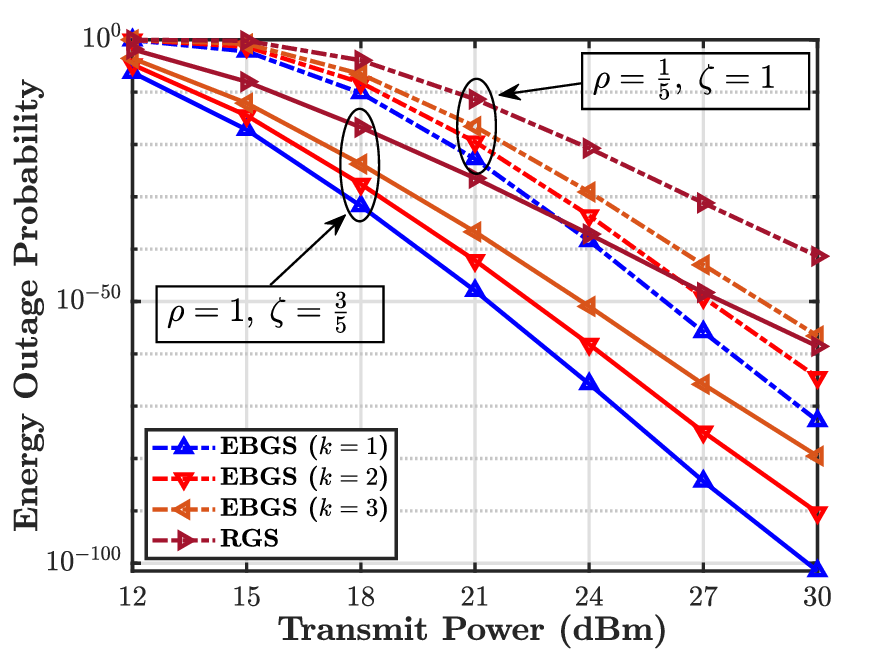}
    \vspace{-2mm}
     \caption{}
    \vspace{-2mm}
    \label{nl_zeta}
\end{subfigure}
\begin{subfigure}[b]{.5\textwidth}
    \centering
    \includegraphics[width=0.62\linewidth]{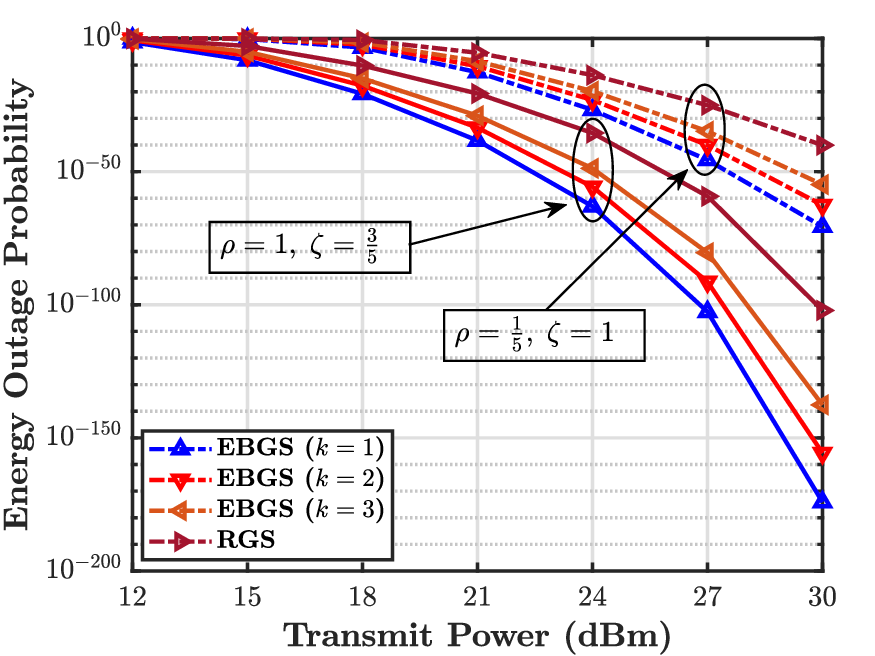}
    \vspace{-2mm}
    \caption{}
    \vspace{-2mm}
    \label{ebgs}
\end{subfigure}
\caption{\footnotesize  Variation of $P_{\rm tx}$ with energy outage in (a) a linear EH model and (b) a nonlinear EH model.}
\vspace{-4mm}
\end{figure*}

Fig. \ref{sbgs} illustrates the impact of the SNR on data outage for the proposed group selection strategies. Specifically, from \eqref{35} and \eqref{39}, we evaluate how the SBGS and RGS schemes perform in terms of data outages for $k$-th best selection for different $k$ values. Note that the necessary energy is harvested according to the linear or nonlinear EH model of the PS or TS configuration, and their effects on the received SNR are preserved. Here, we consider a simulation setup of each group consisting of $10$ patches and $10$ groups are available to aid communication. We observe that the outage probability for all the cases of SBGS and RGS exhibits a decreasing trend with increasing SNR values. This is due to the fact that rising SNR raises the achievable data rate, which causes the outage probability to decrease. Moreover, we observe that the outage performance of RGS is inferior to SBGS implying that a strategic selection is always better than the random selection.  Furthermore, the performance of the $k$-th best group selection is superior to the $(k+1)$-th best group selection. This is because, in this case, we are choosing the optimal group as compared to having a sub-optimal choice.

Fig. \ref{snr_comp} demonstrates the role of spatial correlation at the RIS on the outage probability by using the inter-patch distance as $\lambda/8$. By considering the system parameters as in Fig. \ref{sbgs},  we compare our method with an existing framework \cite{channelestimation} that groups the RISs without taking spatial correlation into account. The figure shows that the curves are in a decreasing trend with respect to the increasing SNR values, and the outage for uncorrelated scenario is less than that of its correlated counterpart. This is because, in an uncorrelated scenario, one channel being unusable for communication does not necessarily affect its neighboring channels, which is not the case in a correlated scenario. This may give us an impression that it is unwise to take spatial correlation into account. However, in the uncorrelated scenario,  the inter-patch distance in an RIS of finite dimension cannot be reduced below $\lambda/2$, which severely limits the number of available patches and thus the data rate. On the other hand, from \eqref{ypsdef} and \eqref{20}, we observe that SNR is a function of $h_c$ and $g_c$, which are the sum of correlated channels. This indicates that with reduced inter-patch distance, more patches are accessible for the correlated case, which eventually raises the data rate. Lastly, as also observed previously, the outage corresponding to the RGS scheme results in the worst outage performance.

\begin{figure}
    \centering
    \includegraphics[width=0.65\linewidth]{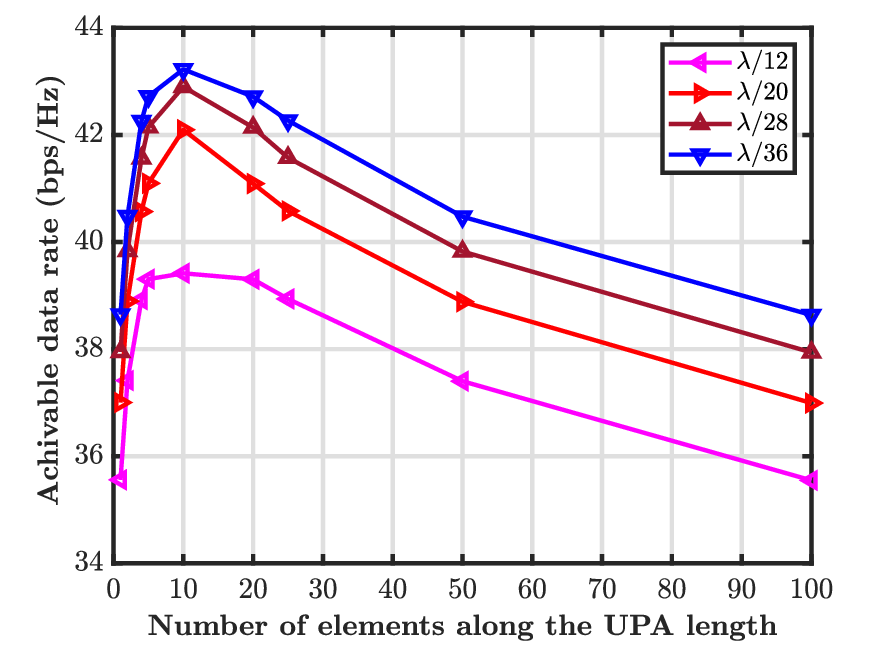}
    \caption{Impact of RIS configuration on achievable data rate.}
    \label{upa}
\end{figure}

The effect of EVT on the SBGS selection scheme, as evident in \eqref{59}, is shown in Fig. \ref{evt}, where we consider $M=10$, Rician factor $K=1$, required SNR threshold $\gamma_{th}=5$ dB, and the number of groups $B=20,80$ and $140$, respectively. In this figure, we perform a $k$-th best selection scheme for $k=1,6$. Note that, regardless of the value of $B$, the outage performance demonstrates a decreasing trend, which has also been previously observed. Moreover, for a given SNR, the outage performance sharply improves with increasing $B$. The reason for this is attributed to the fact that as $B$ increases, the `chance of the best group being available' also increases, which improves the outage performance.

Fig. \ref{nl_zeta} and Fig. \ref{ebgs} show the impact of transmit power on the energy outage as computed in \eqref{49} for the $k$-th best group selection in a linear and non-linear EH scenario, respectively. 
Here, we consider both the PS and TS configurations by using  $\left(\zeta=1,0<\rho<1\right)$ and $\left(\rho=1,0<\zeta<1\right)$, respectively.
We consider that each group has $10$ patches and $40$ groups are accessible to facilitate communication, distance from $S$ to $R_i$ is $15$ m, $P_{ph}=5$ dBm, and $P_t=5$ dBm \cite{Tnse}. In both the figures, we observe that, irrespective of $k$, the energy outage exhibits a decreasing trend with increasing transmit power, which is intuitive. This is supported by the fact that the suggested selection schemes perform better as the transmit power increases. Moreover, we notice that the energy outage for the $1$st best group is lesser than the $2$nd best, which in turn, is lesser than the $3$rd best. Consequently, as also observed earlier, we find that the RGS scheme performs the worst.

Fig. \ref{upa} illustrates the impact of RIS configuration on the achievable data rate for various inter-patch distance; here we consider the Rician $K=2$, power splitting factor $\rho=0.2$, and the number of elements in each group $M=100$. Specifically, for a fixed inter-patch distance, we arrange these $M$ elements in various uniform planar array (UPA)-based configurations. Therefore, the configuration $(N_x,N_y) \in \{ (1,100),(2,50),$ $(4,25),(5,20),(10,10),(20,5),(25,4),(50,2),(100,1)  \}$, where $N_x$ and $N_y$ represents the number of elements along the UPA length and width, respectively. We observe that, irrespective of the UPA configuration, the data rate increases with decreasing inter-patch distance. Moreover, with all other system parameters remaining constant, the maximum data rate is achieved for the square configuration, i.e., $N_x=N_y=10$. The reason for this is attributed to the fact that, among all the possible configurations, $N_x=N_y$ results in the most compact shape with minimum distance among any pair of farthest located reflecting elements in the group. Hence, in scenarios, where it is not possible to have $N_x=N_y$, the best result is obtained by choosing $N_x,N_y$ very close to each other.


\section{Conclusion}\label{conclu}
In this paper, we proposed novel order statistics-based generalized group selection strategies for a self-sustainable RIS-assisted D2D communication set-up in a Rician fading scenario, which takes into account the aspect of correlated channels. These strategies are based on the performance metrics like the end-to-end SNR and the energy harvested at the RIS group. Here, we considered both the PS and the TS configurations of the RIS and provided performance bounds for the system parameters of concern. We demonstrated the importance of spatial correlation at the RIS (in terms of the inter-patch spacing) on the system performance by analytically characterizing both the data outage and the energy outage. Extensive Monte Carlo simulations validate the proposed framework. Moreover, by using EVT, we further looked into the asymptotic scenario of having a large number of groups available at the RIS for selection. Thereafter, we observed that the outage performance, in this case, monotonically improves with increasing number of groups. As an immediate extension of this work, we intend to investigate the aspect of considering an adaptive selection of number of elements in the group for self-sustainable RIS-aided communication scenario. Finally, the scenario of incorporating the aspect of multi-antenna source/destination, phase noise at the RIS, and distributed multi-RIS configurations can also be considered.


\section*{Appendices}
\addcontentsline{toc}{section}{Appendices}

\renewcommand{\thesubsection}{\Alph{subsection}.}

\subsection{Proof of Theorem \ref{theo_1}}  \label{thrm1}
As $\rho$ is equal for all the $M$ reflecting elements of $R_i$, from \eqref{E_L}, we obtain the harvested energy
\begin{equation}\label{abc1}
    E_{l,PS}=T_s \rho P_{\rm tx} \rho_Ld_{S,R_i}^{-\alpha}\sum_{i=1}^{M}|{ \Tilde{h_i}}|^2.
\end{equation}
By rewriting \eqref{abc1}, we obtain 
\begin{equation}\label{p-tx}
    P_{\rm tx}=\frac{E_{l,PS}}{T_s \rho \rho_Ld_{S,R_i}^{-\alpha}\sum_{i=1}^{M}|{ \Tilde{h_i}}|^2}.
\end{equation}
Accordingly, the received SNR $\gamma_{\rm PS}$ at $D$ is
\begin{align}  \label{12}
    \gamma_{\rm PS} &= \frac{\big(1-\rho\big) P_{\rm tx} \rho^2_L\big(d_{S,R_i}d_{R_i,D}\big)^{-\alpha}}{\sigma^2_0}|g_c|^2|h_c|^2.
\end{align}
By using \eqref{p-tx} in \eqref{12}, we have
    \begin{align}  \label{gps1}
    \gamma_{\rm PS}&= \left(\frac{1}{\rho}-1\right) \frac{\rho_L E_{l,PS} \times d_{R_i,D}^{-\alpha}}{T_s \sigma^2_0 \sum_{i=1}^{M}|{\Tilde{h_i}}|^2 }|g_c|^2|h_c|^2
\end{align}
and the resultant achievable data rate is
\begin{equation}  \label{rps}
    R_{\rm PS}=\log_2\big(1+\gamma_{\rm PS}\big).
\end{equation}
Here, we observe that the power splitting factor $\rho$ has a significant impact on both the EH and IT performances. Therefore, we now investigate the performance bounds on $\rho$, which will allow the PS configuration to function flawlessly.

According to Section \ref{psdef}, the $R_i$ will not be able to transfer the incoming signal in the desired direction if $E_{l, PS}<E_{\rm req,PS}$. As a result, for proper functioning of $R_i$, we require
\begin{equation}\label{18}
E_{l, PS} \geq E_{\rm req,PS}=T_s\Big(MP_{t}+P_{ph} \Big) .
    \end{equation}
Hence, by using \eqref{E_req_PS} and \eqref{abc1}, we obtain 
\begin{equation}  
    T_s \rho P_{\rm tx} \rho_Ld_{S,R_i}^{-\alpha}\sum_{i=1}^{M}|{ \Tilde{h_i}}|^2 \geq T_s\Big(MP_{t}+P_{ph} \Big),
    \end{equation}
which finally results in    
    \begin{equation}  \label{l_ps1}
    \frac{MP_{t}+P_{ph}}{ P_{\rm tx} \rho_Ld_{S,R_i}^{-\alpha}\sum_{i=1}^{M}|{ \Tilde{h_i}}|^2} \leq \rho.
\end{equation}
Therefore, if the application-specific minimum required data rate is $R_{\rm req}$, we must have
\begin{equation}\label{rho_up}
    R_{\rm req}\leq R_{\rm PS}=\log_2{(1+\gamma_{\rm PS})}.
\end{equation}
As a result, from \eqref{gps1} and \eqref{rho_up}, we obtain
\begin{align}
    & 2^{R_{\rm req}} -1 \leq \gamma_{\rm PS} =\left(\frac{1}{\rho}-1\right)  \eta,
\end{align}
where $\displaystyle \eta= \frac{\rho_L E_{l,PS} \times d_{R_i,D}^{-\alpha}}{T_s \sigma^2_0 \sum_{i=1}^{M}|{\Tilde{h_i}}|^2 }|g_c|^2|h_c|^2$. This, after trivial algebraic manipulations, results in
\begin{align}  \label{l_ps2}
    &\frac{2^{R_{\rm req}} -1}{\eta}\leq \frac{1}{\rho}-1\implies \rho \leq \frac{\eta}{2^{R_{\rm req}} -1+\eta}.
\end{align}
Hence, by combining \eqref{l_ps1} and \eqref{l_ps2}, we obtain \eqref{theolin1}.
\subsection{Proof of Theorem \ref{th_2}}\label{appen2}

Note that, for the proper functioning of $R_i$, we require $E_{nl, PS} \geq E_{\rm req, PS}$, i.e., we must have
\begin{equation}
    E_{nl, PS}^{\max} \geq E_{\rm req,PS} \:\: {\rm and} \:\: E_{nl, PS}^{\min} \geq E_{\rm req,PS}.
\end{equation}
Hence, if we consider the best-case scenario, i.e., $|\Tilde{h}_i|=|h_{\rm \max}|$ $\forall$ $i=1,\cdots,M$, by using \eqref{E_req_PS} and \eqref{E_No1n}, we obtain 
\begin{align}
    & MT_s\Big(\frac{a\rho P_{\rm tx} \rho_L(d_{S,R_i})^{-\alpha}|h_{\rm \max}|^2+b}{\rho P_{\rm tx} \rho_L(d_{S,R_i})^{-\alpha}|h_{\rm \max}|^2+c}-\frac{b}{c}\Big) \geq E_{\rm req,PS}\nonumber\\
    & \implies \!\!\!\Big(\frac{a\rho P_{\rm tx} \rho_L(d_{S,R_i})^{-\alpha}|h_{\rm \max}|^2+b}{\rho P_{\rm tx} \rho_L(d_{S,R_i})^{-\alpha}|h_{\rm \max}|^2+c}-\frac{b}{c}\Big) \geq \frac{MP_{t}+P_{ph}}{M}.
    \end{align}
On further simplification, we obtain
\begin{equation}  \label{n_up}
    \frac{c(MP_{t}+P_{ph})}{M\rho_L(d_{S,R_i})^{-\alpha} P_{\rm tx} |h_{\rm \max}|^2\Big(a-\frac{MP_{t}+P_{ph}}{M}-\frac{b}{c}\Big)}  \leq \rho.
\end{equation}
Similarly, now we consider the worst-case scenario, i.e., $|\Tilde{h}_i|=|h_{\rm \min}|$ $\forall$ $i=1,\cdots,M$ and from \eqref{E_No1n}, we have
\begin{equation}
    P_{\rm tx}=\frac{\frac{cE_{nl,PS}^{\min}}{MT_s}}{\rho \rho_L(d_{S,R_i})^{-\alpha} |h_{\min}|^2\left(a-\frac{E_{nl,PS}^{\min}}{MT_s}-\frac{b}{c}\right )}.
\end{equation}
Accordingly, the received SNR $\gamma_{\rm PS}$ at $D$ is
\begin{align}  \label{12a}
    \gamma_{\rm PS} &= \frac{\big(1-\rho\big) P_{\rm tx} \rho^2_L\big(d_{S,R_i}d_{R_i,D}\big)^{-\alpha}}{\sigma^2_0}|g_c|^2|h_c|^2\nonumber\\
    &=\left(\frac{1}{\rho}-1\right)\frac{cE_{nl,PS}^{\min}\rho_L d_{R_i,D}^{-\alpha}}{MT_s |h_{\min}|^2\left(a-\frac{E_{nl,PS}^{\min}}{MT_s}-\frac{b}{c}\right ) \sigma^2_0 } |g_c|^2|h_c|^2.
\end{align}
Note that, irrespective of the channel conditions, the data rate needs to be above the threshold $R_{\rm req}$, i.e., \eqref{rho_up} results in
\begin{align}
   R_{\rm req}& \leq R_{\rm PS} \implies 2^{R_{\rm req}}-1 \leq \gamma_{PS}=\left(\frac{1}{\rho}-1\right)\kappa,
\end{align}
where
\begin{equation}
    \kappa=\frac{cE_{nl,PS}^{\min}\rho_L d_{R_i,D}^{-\alpha}}{MT_s |h_{\min}|^2\left(a-\frac{E_{nl,PS}^{\min}}{MT_s}-\frac{b}{c}\right ) \sigma^2_0 } |g_c|^2|h_c|^2.
\end{equation}
Thereafter, further simplifications result in
\begin{equation}\label{dnl}
    \rho \leq \frac{\kappa}{(2^{R_{\rm req}}-1)+\kappa}.
\end{equation}
Finally, by combining \eqref{n_up} and \eqref{dnl}, we have \eqref{t2}.

\subsection{Proof of Theorem \ref{theo_3}}\label{appen3}
The harvested energy, in a slot, is obtained from \eqref{E_L} as
\vspace{-2mm}
\begin{equation}\label{abc}
    E_{l,TS}=\zeta T_s  P_{\rm tx} \rho_Ld_{S,R_i}^{-\alpha}\sum_{i=1}^{M}|{ \Tilde{h_i}}|^2.
\end{equation}
For proper RIS functioning, we must have $E_{l,TS} \geq E_{\rm req,TS}$, i.e., by using \eqref{TS_ener} and \eqref{abc}, we get
\begin{align}\label{34}
    & E_{l,TS} \geq E_{\rm req,TS}=T_s\Big(MP_{t}(1-\zeta)+P_{ph} \Big) \nonumber\\
    & \implies  \zeta T_s P_{\rm tx} \rho_Ld_{S,R_i}^{-\alpha}\sum_{i=1}^{M}|{ \Tilde h_i}|^2 \geq T_s\Big(MP_{t}(1-\zeta)+P_{ph} \Big) \nonumber \\
    & \implies  \zeta P_{\rm tx} \rho_Ld_{S,R_i}^{-\alpha}\sum_{i=1}^{M}|{ \Tilde h_i}|^2 +\zeta MP_{t} \geq \Big(MP_{t}+P_{ph} \Big), \nonumber
    \end{align}
    which finally results in
\begin{equation} \label{d_ts2}
      \frac{MP_{t}+P_{ph}}{ MP_t+P_{\rm tx} \rho_Ld_{S,R_i}^{-\alpha}\sum_{i=1}^{M}{|\Tilde h_i}|^2} \leq \zeta.
\end{equation}
Similar to the PS configuration, if the application-specific minimum required data rate is $R_{\rm req}$, we have
 \begin{align}
      R_{\rm req}\leq R_{\rm TS} &=  (1-\zeta)\log_2({1+\gamma_{\rm TS}})
      =(1-\zeta) R_{\rm arc},
 \end{align}
 where $R_{\rm arc}=\log_2({1+\gamma_{\rm TS}})$ is the achievable data rate and
 \begin{equation}
    \gamma_{\rm TS} = \frac{ P_{\rm tx} \rho^2_L\big(d_{S,R_i}d_{R_i,D}\big)^{-\alpha}}{\sigma^2_0}|g_c|^2|h_c|^2.
  \label{gts}\end{equation}
Further simplification of the above results in 
      \begin{align}
     & \zeta \leq 1-\frac{R_{\rm req}}{R_{\rm arc}}.\label{d_Ts1}
 \end{align}
Thus, by combining \eqref{d_ts2} and \eqref{d_Ts1}, we obtain \eqref{th1_3}.

\subsection{Proof of Theorem \ref{theo_4}}\label{appen4}
If $E_{nl,TS}^{\max}$ is the harvested energy by considering $|\Tilde{h_i}|=|h_{\max}|\; \forall \;i=1,\cdots, M$, from the proper functioning condition of $R_i$, we have, $E_{nl,TS}^{\max}\geq E_{\rm req, TS}$. Hence, by using \eqref{TS_ener} and \eqref{E_Non}, we obtain
\begin{equation}
      \zeta T_sM\Big(\frac{a\rho P_{\rm tx} \rho_L(d_{S,R_i})^{-\alpha}|h_{\rm \max}|^2+b}{\rho P_{\rm tx} \rho_L(d_{S,R_i})^{-\alpha}|h_{\rm \max}|^2+c}-\frac{b}{c}\Big) \geq E_{\rm req,TS}.
    \end{equation}
The above equation can be rewritten as
    \begin{align}
      & \zeta M\left(\frac{a \varphi+b}{ \varphi+c}-\frac{b}{c}\right ) \geq (1-\zeta)MP_t+P_{ph}\nonumber\\
     & \implies \zeta M\left(\frac{a \varphi+b}{ \varphi+c}-\frac{b}{c}\right )+\zeta MP_t \geq MP_t+P_{ph}
\end{align}
where $\varphi=\rho_L d_{S,R_i}^{-\alpha} P_{\rm tx}|h_{\max}|^2$. Thereafter, further simplification results in
\begin{align}\label{tsnz}
    & \frac{MP_t+P_{ph}}{M\left(P_t+\frac{a \varphi+b}{ \varphi+c}-\frac{b}{c}\right )}\leq \zeta.
\end{align}
Note that, during the EH process, we are concerned only with the $S-R_i$ link; hence, the role of $g_c$ is irrelevant. Moreover, from \eqref{com_h}, we have $h_c=\sum\limits_{i=1}^{M}\Tilde{h_i},$
which results in $|h_c|^2=M^2|h_{\min}|^2$ for the considered worst-case scenario. Thus, the resultant SNR at $D$  is
\begin{equation}
    \gamma_{\rm TS}^{\min} = \frac{ M^2P_{\rm tx} \rho^2_L\big(d_{S,R_i}d_{R_i,D}\big)^{-\alpha}}{\sigma^2_0}|g_c|^2|h_{\min}|^2
\end{equation}
and the resultant achievable data rate is
\begin{equation}
    R_{\rm TS}=\big(1-\zeta \big)\log_2(1+\gamma_{\rm TS}^{\min}).
\end{equation}
Regardless of the channel conditions, the data rate must exceed the specified threshold $R_{\rm req}$, i.e.,
\begin{align}
      R_{\rm req}\leq R_{\rm TS} &=  (1-\zeta)\log_2({1+\gamma_{\rm TS}^{\min}})
      =(1-\zeta) R_{\rm arc}^{\min},\nonumber
 \end{align}
where we define $R_{\rm arc}^{\min}=\log_2({1+\gamma_{\rm TS}^{\min}})$. Further simplification of the above results in
      \begin{align}
     & \zeta \leq 1-\frac{R_{\rm req}}{R_{\rm arc}^{\min}}.\label{d_Ts12}
 \end{align}
Hence, by combining \eqref{tsnz} and \eqref{d_Ts12}, we obtain \eqref{ze_n}.

\subsection{Proof of Theorem \ref{th_5}}  \label{theo1}
From \eqref{outage3}, we have $\mathcal{O}^{RGS}_{\rm PS/TS}=\mathbb{P}\left( \gamma_{\rm PS/TS} \leq 2^{\frac{R_{\rm req}}{f(\zeta)}}-1 \right )$.
Hence, by using \eqref{12} and \eqref{fzdef}, we obtain
\begin{align}
  &\mathcal{O}^{RGS}_{\rm PS} \nonumber \\
  &= \mathbb{P} \left( \frac{\big(1-\rho\big) P_{\rm tx} \rho^2_L\big(d_{S,R_i}d_{R_i,D}\big)^{-\alpha}}{\sigma^2_0}|g_c|^2|h_c|^2 \leq 2^{R_{\rm req}}-1 \right) \nonumber \\
  &= \mathbb{P} \left( |g_c|^2|h_c|^2 \leq \frac{2^{R_{\rm req}}-1}{\big(1-\rho\big)\Psi} \right)\label{out_ran},
\end{align}
where $\displaystyle \Psi=\frac{P_{\rm tx} \rho^2_L\big(d_{S,R_i}d_{R_i,D}\big)^{-\alpha}}{\sigma^2_0}$. From \eqref{jpdf}, we know that the quantity $Z=|g_c|^2|h_c|^2$ is approximated by a Gamma random variable. Therefore, by using its CDF, we have
\begin{align}
    \mathcal{O}^{RGS}_{\rm PS} &= \int_{0}^{\frac{2^{R_{\rm req}}-1}{\big(1-\rho\big)\Psi}}f_Z(y)dy=\frac{\gamma\left(\alpha, \frac{2^{R_{\rm req}}-1}{\delta \big(1-\rho\big)\Psi}\right)}{\Gamma(\alpha)}.
\end{align}
Similarly, by using \eqref{gts} and \eqref{outage3}, we obtain
\begin{align}
\mathcal{O}^{RGS}_{\rm TS} &= \mathbb{P} \left( |g_c|^2|h_c|^2 \leq \frac{2^{\frac{R_{\rm req}}{1-\zeta}}-1}{\Psi} \right) 
=\frac{\gamma\left(\alpha, \frac{2^{\frac{R_{\rm req}}{1-\zeta}}-1}{\delta \Psi}\right)}{\Gamma(\alpha)}\nonumber.
\end{align}
\subsection{Proof of Theorem \ref{th_6}} \label{theo2}

In Section \ref{ana_channel}, we approximate $|g_c|^2|h_c|^2$ by a Gamma distribution. Therefore, from \eqref{out_ran}, the SNR $\big(1-\rho\big)\Psi|g_c|^2|h_c|^2$ is a scaled Gamma random variable.
Hence, the outage probability for the SBGS scheme, in the PS configuration, is
\begin{equation}\label{outage5}
    \mathbb{P}\left(\log_2(1+X_{i^*})<R_{\rm req}\right)=\mathbb{P}\left(X_{i^*}\leq x\right),
\end{equation}
where $x=2^{R_{req}}-1$. Also, from \eqref{pdf_or} and \eqref{outage5}, the CDF of $X_{i^*}$  is obtained as
 \begin{align}\label{out_sn}
    &\mathcal{O}^{SBGS}_{k,\rm PS}=\int_{0}^{x}f_{X_{i^*}}(z)dz\nonumber\\
    & =k\binom{|\mathcal{S}|}{k}\int_{0}^{x}f_{X_i}(z)F_{X_i}(z)^{|\mathcal{S}|-k}(1-F_{X_i}(z))^{k-1}dz
\end{align}
In \eqref{out_sn}, by substituting $y=F_{X_i}(x)$ and using the definition of normalized incomplete Beta function \cite{grad}, we obtain 
\begin{equation}\label{PS_snr}
    \mathcal{O}^{SBGS}_{k,\rm PS}=\mathcal{I}_{F_{X_i}(x)}\left(|\mathcal{S}|-k+1,k\right).
\end{equation}
Similarly, now we obtain the outage of the $k$-th best group selection in the TS configuration. Note that in TS configuration, the outage is obtained by \eqref{PS_snr} itself, where the only difference is that in this case, we have $\rho=0$ and $x=2^{\frac{R_{\rm req}}{1-\zeta}}-1$.

\subsection{Proof of Theorem \ref{th_7}} \label{theo3}

From \eqref{18}, we observe that a group is able to transfer information if it harvests the minimum required energy.
Therefore, for the PS configuration, the outage of the $k$-th best group in the EBGS scheme is  
\begin{equation}\label{81}
    \mathbb{P}\left(E_{i^*}\leq E_{\rm req,PS}\right),
\end{equation}
where $E_{\rm req,PS}$ is the minimum required energy in PS configuration. Hence, from \eqref{pdf_en}, the outage of the $k$-th best group is obtained as
\begin{align}
    \mathbb{O}^{EBGS}_{\rm PS/TS}&=\int_{0}^{x}f_{E_{i^*}}(z)dz\nonumber\\
    & =k\binom{|\mathcal{A}|}{k}\int_{0}^{x}f_{E_i}(z)F_{E_i}(z)^{|\mathcal{A}|-k}(1-F_{E_i}(z))^{k-1}dz\nonumber\\
    & =\mathcal{I}_{F_{E_i}(x)}\left(|\mathcal{A}|-k+1,k\right),
\end{align}
where $f_{E_{i^*}}(z)$ denotes the PDF of the $k$-th best group.

For the linear EH model, by using \eqref{E_L} we have $ E_{i}=\nu\sum\limits_{j=1}^{M}|{ \Tilde{h_j}}|^2,$ where $\nu=T_s \rho P_{\rm tx} \rho_Ld_{S,R_i}^{-\alpha}.$ Therefore, with all the other parameters remaining constant, $E_{i}$ is a function of $\sum\limits_{j=1}^{M}|{ \Tilde{h_j}}|^2$. Since $|{ \Tilde{h_j}}|^2$ follows the non-central $\chi^2$ distribution, we approximate $E_i$ as a Gamma random variable\cite{sum_non_chi}. Now, for the non-linear EH model, by using \eqref{E_Non}, which followed by trivial algebraic manipulations, results in
\begin{align}\label{simp}
    \frac{aP_{\rm tx}|h|^2+b}{P_{\rm tx}|h|^2+c}-\frac{b}{c}
    &=\frac{ac-b}{c}+(b-ac)\times \frac{1}{P_{\rm tx}|h|^2+c}.
\end{align}
Accordingly, by using \eqref{simp}, the harvested energy is 
\begin{align}
E_i &=T_s\sum_{j=1}^{M}\Big(\frac{a P_{\rm tx} \rho_L(d_{S,R_i})^{-\alpha}|\Tilde{h_j}|^2+b}{ P_{\rm tx} \rho_L(d_{S,R_i})^{-\alpha}|\Tilde{h_j}|^2+c}-\frac{b}{c}\Big)\nonumber\\
    &=T_s\left(\frac{M(ac-b)}{c}+\sum_{j=1}^{M} \frac{b-ac}{\rho_L(d_{S,R_i})^{-\alpha}P_{\rm tx}|\Tilde{h_j}|^2+c}\right). \label{nl_ene}
\end{align}
Here, \eqref{simp} can be approximated to an inverse Gamma approximation \cite{inverse_gamma}. Therefore, \eqref{nl_ene} can also be approximated to a scaled inverse Gamma random variable.

Lastly, note that if we set $\rho$ as unity in \eqref{E_L}, and substitute $E_{\rm req,PS}$ by $E_{\rm req,TS}$ in \eqref{81}, in a similar way, we obtain the outage performance corresponding to the TS configuration for both the linear and non-linear cases.

\bibliographystyle{IEEEtran}
\bibliography{ref}

\end{document}